%% file: DeFog-v1.tex
\documentclass[sigconf]{acmart}

\usepackage{url}
\usepackage{footnote}
\makesavenoteenv{tabular}
\makesavenoteenv{table}
\usepackage{graphicx}
\usepackage{subfig}
\usepackage{todonotes}
\usepackage{xcolor}
\usepackage{booktabs}
\usepackage{multirow, makecell}

\def\BibTeX{{\rm B\kern-.05em{\sc i\kern-.025em b}\kern-.08emT\kern-.1667em\lower.7ex\hbox{E}\kern-.125emX}}
    
\copyrightyear{2019}
\acmYear{2019}
\setcopyright{acmlicensed}
\acmConference[SEC '19]{SEC '19: ACM/IEEE Symposium on Edge Computing}{November 07--09, 2019}{Washington, DC}
\acmBooktitle{SEC '19: ACM/IEEE Symposium on Edge Computing, November 07--09, 2019, Washington, DC}
\acmPrice{15.00}
\acmDOI{10.1145/1122445.1122456}
\acmISBN{978-1-4503-9999-9/18/06}

\begin{document}

%
\title[\texttt{DeFog}: Fog Computing Benchmarks]{\texttt{DeFog}: Fog Computing Benchmarks}

\author{Jonathan McChesney}
\affiliation{%
  \institution{Queen's University Belfast, UK}
}
\email{jmcchesney01@qub.ac.uk}

\author{Nan Wang}
\affiliation{%
  \institution{Queen's University Belfast, UK}
  }
\email{nwang03@qub.ac.uk}

\author{Ashish Tanwer}
\affiliation{%
  \institution{Cisco Systems, USA}
  }
\email{astanwer@cisco.com}

\author{Eyal de Lara}
\affiliation{%
  \institution{University of Toronto, Canada}
  }
\email{delara@cs.toronto.edu}

\author{Blesson Varghese}
\affiliation{%
  \institution{Queen's University Belfast, UK}
  }
\email{b.varghese@qub.ac.uk}

\renewcommand{\shortauthors}{McChesney, et al.}

\begin{abstract}
\input{abstract.tex}
\end{abstract}

%
%



\settopmatter{printfolios=true}

\maketitle

\section{Introduction}
\label{sec:introduction}
\input{introduction}

\section{DeFog Benchmarking}
\label{sec:defogbenchmarking}
\input{defog}

\section{Benchmark Applications and Metrics Collected by \texttt{DeFog}}
\label{sec:applicationsandmetrics}
\input{applicationsandmetrics}

\section{Experimental Studies}
\label{sec:experimentalstudies}
\input{experiments}

\section{Related Work}
\label{sec:relatedwork}
\input{relatedwork}

\section{Conclusions}
\label{sec:conclusions}
\input{conclusions.tex}

\balance
\bibliographystyle{ACM-Reference-Format}
\bibliography{references}

\end{document}

%% file: abstract.tex
Fog computing envisions that deploying services of an application across resources in the cloud and those located at the edge of the network may improve the overall performance of the application when compared to running the application on the cloud. However, there are currently no benchmarks that can directly compare the performance of the application across the cloud-only, edge-only and cloud-edge deployment platform to obtain any insight on performance improvement. 
This paper proposes \texttt{DeFog}, a first Fog benchmarking suite to: (i) alleviate the burden of Fog benchmarking by using a standard methodology, and (ii) facilitate the understanding of the target platform by collecting a catalogue of relevant metrics for a set of benchmarks. The current portfolio of \texttt{DeFog} benchmarks comprises six relevant applications conducive to using the edge. Experimental studies are carried out on multiple target platforms to demonstrate the use of \texttt{DeFog} for collecting metrics related to application latencies (communication and computation), for understanding the impact of stress and concurrent users on application latencies, and for understanding the performance of deploying different combination of services of an application across the cloud and edge. \texttt{DeFog} is available for public download (\url{https://github.com/qub-blesson/DeFog}).

%% file: introduction.tex
The premise of the emerging Fog computing model is to bring appropriate services of an application from the cloud to the edge of the network~\cite{edgecomputing-00, fogcomputing-00}. Since the edge is closer to end user-devices, running services on edge resources will reduce the communication latencies of applications by enabling more data to be processed near the end user-device before it is sent to (or processed on) the cloud~\cite{edgecomputing-01, edgecomputing-03, shi-edgecomputing}. Consequently, the overall Quality-of-Service (QoS) and Quality-of-Experience (QoE) of an application can be improved~\cite{edgeapplication-01}. 

Given that Fog computing is in its nascent stages of development, there are a number of challenges that need to be addressed.
{\color{black}One such challenge is understanding the relative performance of Fog applications by comparing target hardware platforms from different vendors due to diversity of hardware architectures and the impact on performance when system software level changes or new networking protocols are introduced.  
Fog benchmarking solutions are required to address this. 
}

Benchmarking is a commonly used technique for evaluating the relative performance benefits of different target computing models (or platforms) and the applications that run on them~\cite{benchmarking-01, benchmarking-02}. Benchmarks capture a variety of workloads that are likely to be executed using a computing model. These workloads are systematically and repeatedly executed so as to obtain a large catalogue of relevant metrics related to performance, which is subsequently analysed for obtaining concrete answers to questions posed by hardware vendors or system software developers. There are dedicated benchmarks that have been developed for supercomputers, namely LINPACK~\cite{linpack-01} and NAS Parallel Benchmarks~\cite{npb-01}, or for the cloud, namely CloudRank-D~\cite{cloudrankd-01} and DCBench~\cite{dcbench-01}. However, we note that there are no such benchmarking solutions available for Fog computing. Therefore, the focus of the research presented in this paper is to report the development of a first Fog benchmarking suite, referred to as \texttt{DeFog} (the rationale for the name is to `demystify cloud-edge interactions of a Fog system').

The \texttt{DeFog} suite is underpinned by a six step benchmarking methodology that operates in three deployment modes, namely a cloud-only mode, edge-only mode, and cloud-edge (Fog) mode. {\color{black}
We anticipate that \texttt{DeFog} can be used to understand at least the following three fundamental questions:
}

\textbf{Q1}: How can the relative performance of using a Fog platform over the cloud-only platform be understood and quantified? 

\textbf{Q2}: If there are multiple services of an application that can be moved to the edge, then how can performance of using the Fog be understood? 

\textbf{Q3}: If there are multiple competing resources available at the edge suited for a specific service, then which resource should be selected for maximising the overall performance gain? 

Benchmarking across different deployment modes will address \textbf{\textit{Q1}}. \texttt{DeFog} is as hardware agnostic as possible (to the extend current container technology supports) in building and deploying the cloud and edge services of the Fog application. The methodology captures a catalogue of performance metrics that are related to the target platform (cloud and edge resources) and applications running on the platform. 
It would also be possible to obtain insight into the best distribution of services for a given class of application across the cloud and the edge, thereby addressing \textbf{\textit{Q2}}. 

Six applications that are Fog relevant workloads, namely a deep learning-based object recognition, a text-to-speech converter, a text-to-audio forced alignment, an online mobile game, an Internet-of-Things (IoT) application, and real-time face detection from video streams, are considered. The applications are latency critical, stream-based and/or bandwidth intensive, thereby making them ideal candidates for use in \texttt{DeFog}. The experimental results presented highlight the relative performance of the benchmarks across the deployment modes on different target platforms. This demonstrates the feasibility of \texttt{DeFog} for addressing \textbf{\textit{Q3}} considered above. 
The \texttt{DeFog} software 
is publicly available\footnote{\url{https://github.com/qub-blesson/DeFog}}.

The contributions of this paper are as follows:
(i) The development of a platform agnostic Fog benchmarking methodology, \texttt{DeFog}, that operates in three deployment modes.
(ii) The evaluation of six application benchmarks that are relevant to Fog computing and making them available in a repository.
(iii) The identification and collection of a catalogue of metrics that capture the properties of the target platform and the application running on it.
(iv) An experimental evaluation of \texttt{DeFog} across cloud resources and multiple single board computers that are used as edge resources.  

The remainder of this paper is organised as follows. Section~\ref{sec:defogbenchmarking} provides an overview of the benchmarking methodology and the deployment modes of \texttt{DeFog}. Section~\ref{sec:applicationsandmetrics} presents \texttt{DeFog}'s current portfolio of six benchmarks that are candidate Fog applications and the catalogue of metrics that are collected by \texttt{DeFog}. 
Section~\ref{sec:experimentalstudies} presents the results obtained from an experimental study. Section~\ref{sec:relatedwork} considers the related work. Section~\ref{sec:conclusions} concludes this paper by presenting future work.

%% file: defog.tex
This section firstly presents a few observations that have led to the design and development of \texttt{DeFog}, followed by the benchmarking methodology that is incorporated in \texttt{DeFog}, and finally the deployment modes of \texttt{DeFog}. The first version of \texttt{DeFog} is available for download from 
(\url{https://github.com/qub-blesson/DeFog}). 

\subsection{Motivation}
Fog benchmarking as a research area is still in its early stages of development. The dependencies between cloud-edge services of an application are not fully known given that there are only a few open source Fog applications available. 
The following five general observations regarding Fog benchmarking have been considered while developing \texttt{DeFog}:

(i) \textit{Fog benchmarking is complex as dependencies between cloud and edge services of a Fog application need to be considered}: In cloud benchmarking, the dependencies of the application are mostly in the cloud. However, the dependencies between cloud and edge services of a Fog application will need to be considered for Fog benchmarking. In other words, the bespoke requirements and dependencies of an application will need to be considered ideally for benchmarking the interactions in a Fog system, which is considered in \texttt{DeFog}. This makes Fog benchmarking more complex than cloud benchmarking. 

(ii) \textit{Fog benchmarks are not readily available}: There is a limited understanding of the real workloads that can benefit the most from using Fog computing. Consequently, there are no open source Fog benchmarks readily available. While the portfolio of benchmarks for traditional computing platforms, such as high-performance clusters or even the cloud are diverse and comprehensive, Fog benchmarks are not yet developed. \texttt{DeFog} attempts to create an early repository of six benchmark applications that can be expanded upon by the community as our understanding of Fog applications evolves and as applications become available. 

(iii) \textit{Fog benchmarking should generate rapid results}: Ideal benchmarking should generate results quickly and in the context of the Fog it is essential given that the edge is a transient environment. Resources located on the edge (routers or gateways made accessible for general purpose computing) are anticipated to have intermittent profiles when compared to dedicated resources in a data center. \texttt{DeFog} is therefore designed to execute in a lightweight manner and under one minute.  

(iv) \textit{Metrics captured during Fog benchmarking must be generalised to a wide variety of workloads}: Many benchmarking solutions are workload specific and therefore generate workload specific metrics on a target platform. Hence, they cannot provide insight into the suitability of the target platform for a different class of workloads. The aim of \texttt{DeFog} is to evolve over time by adding more workloads so that a wide range of metrics can be captured for diverse Fog platforms and application benchmarks. 

(v) \textit{Benchmarking needs to be consistent}: To ensure that benchmarking is consistent each execution of the benchmark must be on the same application build version and package dependencies. This is ensured in \texttt{DeFog} so that cloud-edge services are consistently benchmarked. In addition, to minimise the impact of any noise within the Fog platform (due to temporary network congestion or spikes in workloads), the benchmark needs to be executed multiple times for a range of input data using identical containers to obtain an averaged values of metrics. 

\subsection{Benchmarking Methodology}
\label{subsec:benchmarkingmethodology}

The aim of \texttt{DeFog} is to automate the deployment of benchmark applications on the target platform, the transfer of assets required for these applications to run on the target platform, the generation of metrics relevant to the target platform and benchmark, and finally gather results. The proposed benchmarking methodology used by \texttt{DeFog} accounts for these and is a sequence of six steps, which is as follows: (1) Build and run the benchmark application container image for the target platforms, (2) transfer the required asset to the cloud resource on the target platform, (3) transfer the required asset to the edge resource on the target platforms, (4) execute the benchmark application, (5) gather the values for the catalogue of metrics, and (6) return the metrics to the user. The third step is required for the edge-only or cloud-edge (Fog) deployment modes. The second to fifth step is repeated multiple times for each application for consistency. The metrics are measured by an observing system. The methodology in detail is as follows:

\textit{Step 1 - Build and run a container}: In this first step, a container image is built and tagged for a target platform (cloud and/or edge) to ensure that a consistent execution environment is available for all future runs of the benchmark. This research employs Docker containers and it is run in the detached mode to allow for concurrent benchmarking tasks to be executed on other application containers. 

\textit{Step 2 - Transfer assets to the cloud}: An asset that is transferred to the cloud is the input data required by the application's service that runs on the cloud. Each benchmark application has its corresponding asset and a single benchmark application may have multiple assets. For example, there are currently six benchmark applications in \texttt{DeFog} and each application has its own assets. In addition, each application may have multiple assets corresponding to similar or different sizes or types of input. This ensures that \texttt{DeFog} can be used for varying inputs of the application to identify performance gain under different operating conditions for the application. Some of the communication related metrics discussed in a subsequent section are also measured during this step.   

\textit{Step 3 - Transfer assets to the edge}: The asset transferred in this case is the input data required by the application's service that will be hosted on the edge resource. Currently, the data required by the application benchmark has to be manually partitioned and provided as assets. Automation within this step is desirable, but is not within the scope of this article. Some of the communication related metrics discussed in a subsequent section are measured during this step.  

\textit{Step 4 - Execute the benchmark application}: Given the running container (from Step 1) and the assets (Step 2 and Step 3) for each application the number of times the benchmark needs to be executed is determined. If there are $N$ assets for an application, then the benchmark is executed $N$ times. The benchmark applications are further considered in Section~\ref{subsec:applications}. Additional benchmarking tools, namely UnixBench\footnote{\url{https://github.com/kdlucas/byte-unixbench/tree/master/UnixBench}} and Sysbench\footnote{\url{https://github.com/akopytov/sysbench}} are included within \texttt{DeFog}. They provide a large number of target platform related metrics, including CPU performance, concurrency and I/O read write.

\textit{Step 5 - Gather the values for a catalogue of identified metrics}: A combination of communication and computation related metrics obtained from DeFog. They can be categorised as: (i) target platform metrics, which are attributes that capture the characteristics of the specific target platform under consideration, (ii) Fog application performance metrics, which are attributes that provide insight into the performance of the application on the target platform, and (iii) metrics obtained from external tools, which are attributes obtained by using third party tools. The metrics will be discussed further in Section~\ref{subsec:metrics}.

\textit{Step 6 - Provide the metrics to the user}:
The target audience that can benefit from \texttt{DeFog}, include (i) vendors of new edge hardware who want to demonstrate the benefit of using the Fog via benchmarks, (ii) Internet Service Providers (ISPs) who want to deploy micro data centres at the edge of the network and want to tabulate performance of Fog applications for their customers, (iii) system software administrators who want to investigate the impact on Fog applications when a system level change, such as update to operating system or libraries, is required on the edge resource, and (iv) network administrators who want to quantify the performance of Fog applications when a new networking protocol is introduced. When the benchmarks have completed execution the metrics are provided to the observing system in the form of both comma separated and verbose text files. 


\begin{figure}[]
	\centering
	\includegraphics[width=0.5\textwidth]{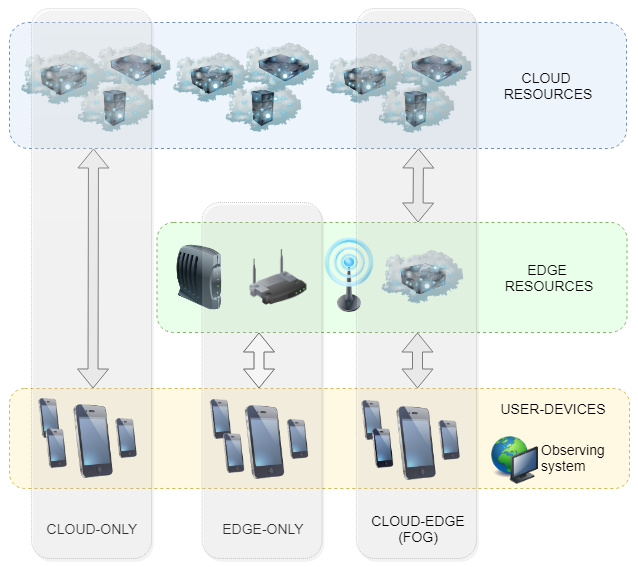}
	\caption{\texttt{DeFog} deployment modes, namely cloud-only, edge-only and cloud-edge (Fog).}
\label{fig:deploymentmodes}
\end{figure}

\subsection{Deployment Modes}
The \texttt{DeFog} benchmarking suite works across three distinct deployment modes, namely a cloud-only, edge-only, and cloud-edge (Fog), as shown in Figure~\ref{fig:deploymentmodes} and considered below. The underlying principle is that the relative performance of the benchmark applications should be compared on different target platforms to quantify any benefit of leveraging the edge for a cloud application. 

The resources available for the target platform as shown in Figure~\ref{fig:deploymentmodes} are from: (i) the cloud resource layer in which a large amount of computational and storage resources are available, (ii) the edge resource layer, which comprises either dedicated micro clouds or traffic routing nodes, such as gateways, routers and base stations, that are augmented with computing and/or storage capabilities, and (iii) the user devices layer, which comprises devices, such as smartphones, wearables, and sensors. A computer system is utilised to observe the benchmarking process and collect the metrics. 

\begin{figure}[ht]
	\centering
	\includegraphics[width=0.5\textwidth]{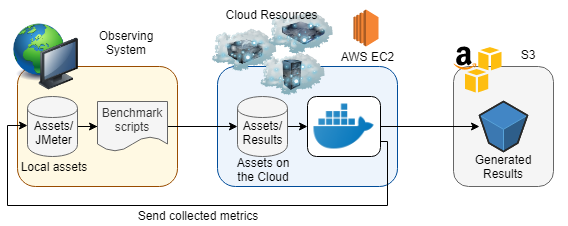}
	\caption{The \texttt{DeFog} cloud-only deployment mode.}
\label{fig:cloudonlydeploymentmode}
\end{figure}

\textbf{\textit{(i) Cloud-only deployment}}: 
The cloud-only deployment is typical of conventional cloud applications in which all requests originating from an end user-device are serviced by a cloud resource. Figure~\ref{fig:cloudonlydeploymentmode} shows the two tier cloud-only deployment mode comprising the cloud resources and user devices. In \texttt{DeFog}, the application is built and deployed on the cloud resource using Docker containers. The application is then executed in the container and the benchmark metrics are generated. The method adopted is based on a container-based cloud benchmarking approach previously reported~\cite{cloudcontainerbenchmarking-01}. The user device uses the secure shell to interact with the cloud container during the benchmarking process.

A \texttt{DeFog} user specifies which applications in the repository need to be benchmarked and the accompanying asset is transferred to the cloud (in this research we used Amazon Web Services (AWS) Elastic Compute Cloud (EC2)\footnote{\url{https://aws.amazon.com/ec2/}}). The output data from the application is uploaded to an AWS Simple Secure Storage (S3)\footnote{\url{https://aws.amazon.com/s3/}} bucket. The output is also transferred to the observing system along with the metrics generated during the benchmarking process. 

\begin{figure}[ht]
	\centering
	\includegraphics[width=0.5\textwidth]{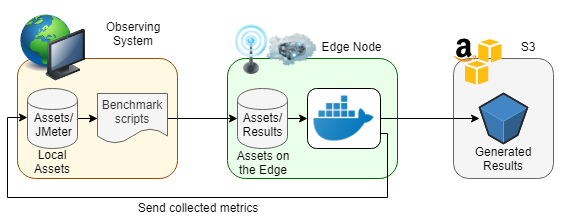}
	\caption{The \texttt{DeFog} edge-only deployment mode.}
\label{fig:edgeonlydeploymentmode}
\end{figure}

\textbf{\textit{(ii) Edge-only deployment}}:
The edge-only deployment mode assumes that all services of an application can be entirely run on an edge resource as shown in Figure~\ref{fig:edgeonlydeploymentmode}. This is practical if it is assumed that dedicated micro clouds or modular data centres are located at the edge of the network and the application provider replicates the application on multiple geographic locations closer to the end user. Similar to the cloud-only deployment mode, the application is deployed on the the edge resource using Docker containers. The application is then run within the container and metrics are generated. In this research resource constrained single board computers, such as Odroid XU4\footnote{\url{https://magazine.odroid.com/odroid-xu4}} and Raspberry Pi 3\footnote{\url{https://www.raspberrypi.org/products/raspberry-pi-3-model-b/}} are used as edge resources. The outputs and the metrics are stored in an S3 bucket as well as sent to the observing system.

\begin{figure}[ht]
	\centering
	\includegraphics[width=0.5\textwidth]{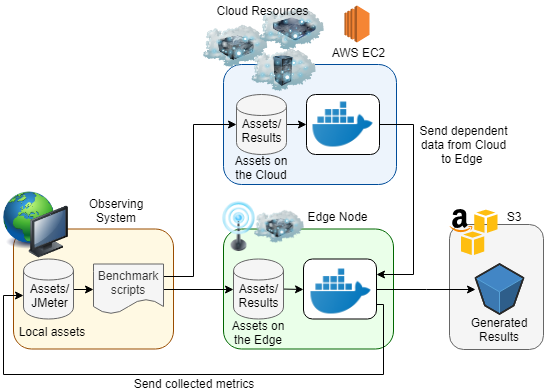}
	\caption{The \texttt{DeFog} cloud-edge (Fog) deployment mode.}
\label{fig:fogdeploymentmode}
\end{figure}

\textbf{\textit{(iii) Cloud-edge (Fog) deployment}}:
In the Fog deployment mode, services of an application may be distributed across the cloud and edge as shown in Figure~\ref{fig:fogdeploymentmode}. Communication latency or bandwidth sensitive services of the application are offloaded to the edge so that the overall QoS of the application can be maximised. 

Consider for example a deep learning application for object detection. This application will require the training of a model, which is a computationally intensive task, and cannot be easily carried out on the edge. If large micro cloud like resources are assumed at the edge, then it may be possible. However this will not be the case if there are only smaller form factor and resource constrained embedded devices available on the edge. Therefore, the training service of the application is ideally suited for the cloud. The detection service could be offloaded to the edge. The assets required by this service are a trained model and associated weights which will need to be offloaded from the cloud to the edge.  

%% file: applicationsandmetrics.tex
\begin{table*}[]
\caption{Fog application benchmarks used in \texttt{DeFog}; Type - Latency Critical (LC), Bandwidth Intensive (BI), Location Aware (LA), Computational Intensive (CI); Single - S, Multiple - M, Multiple, simulated - M(S), Multiple possible, not presented M(N)}
\label{tab:defogapplications}
\begin{tabular}{l|p{5.2cm}|c|c|c|c|c|p{3.5cm}}
\hline
\multirow{2}{*}{\textbf{Application}}                                  & \multirow{2}{*}{\textbf{Description}}             & \multirow{2}{*}{\textbf{Type}} & \multirow{2}{*}{\textbf{\shortstack{End\\device}}} & \multicolumn{2}{c|}{\textbf{Destination}} & \multirow{2}{*}{\textbf{\shortstack{Edge\\services}}} & \multirow{2}{*}{\textbf{\shortstack{Asset transferred from\\cloud to edge for offload}}} \\ \cline{5-6} 
                                                                       &                                                   &                                &                                      & \textbf{Edge}       & \textbf{Cloud}      &                                            &                                                                                 \\ \hline \hline
YOLOv3\footnote{\url{https://github.com/AlexeyAB/darknet}}             & Object classification using deep learning        & BI, CI                         & S, M(S)                              & S                   & S                   & S                                          & Trained model and weights                                                   \\ 
PocketSphinx\footnote{\url{https://github.com/cmusphinx/pocketsphinx}} & Speech-to-text conversion                         & BI, CI                         & S, M(S)                              & S                   & S                   & S                                          & Trained acoustic model                                                      \\ 
Aeneas\footnote{\url{https://github.com/readbeyond/aeneas}}            & Text-audio forced alignment    & BI                             & S, M(S)                              & S                   & S                   & S                                          & Text segment                                                                    \\ 
iPokeMon\footnote{\url{https://github.com/qub-blesson/ENORM}}          & Geo-location based online mobile game             & LC, LA                         & S, M(S)                              & S                   & S                   & S                                          & Location specific data                                                          \\ 
FogLAMP\footnote{\url{https://github.com/foglamp/FogLAMP}}             & IoT edge gateway application & LC                             & M                                    & M                   & M                   & M(N)                                       & NA                                                                              \\ 
RealFD\footnote{\url{https://github.com/qub-blesson/DYVERSE}}          & Real-time face detection on video streams       & LC, BI, CI                     & S                                    & S                   & S                   & M                                          & Depends on edge services                                                     \\ \hline
\end{tabular}
\end{table*}

In this section, six applications that are currently used as benchmarks in \texttt{DeFog} are presented (summarised in Table~\ref{tab:defogapplications}). 
The table shows the coverage of the benchmarks currently available. Latency critical, bandwidth intensive, location aware and computational intensive workloads are represented. In addition, applications with single and multiple users (in many cases multiple users are simulated) are considered. Most applications only use a single edge or cloud resource. However, one application which uses multiple edge and/or cloud destinations is presented. Applications that can offload multiple services to the edge are considered. In addition, the asset required to be transferred from the cloud to the edge for successfully executing an offloaded service is presented. 
Finally, the metrics collected using \texttt{DeFog} are highlighted. 

\subsection{Benchmark Applications}
\label{subsec:applications}

\input{applications}

\subsection{Metrics Collected by Benchmarking}
\label{subsec:metrics}
\input{metrics}

\subsubsection*{Adding New Benchmarks and Metrics}
Extensibility is a consideration in the design of \texttt{DeFog}. Therefore, new benchmarks can be included to the portfolio of applications. 
The benchmarks will need to follow a specific directory structure. The template scripts are provided within the software for setting up and running the containers for the new application. Similarly, new application metrics can be obtained by including appropriate functions in a \texttt{DeFog} script. This is beyond the scope of this paper (detailed steps are provided in the software repository).

%% file: applications.tex
The original version of the applications had to be modified to suit the deployment modes reported in the previous section. The modified version of the applications used in \texttt{DeFog} is available in the project repository\footnote{\url{https://github.com/qub-blesson/DeFog}}. Additional workloads can be added to this portfolio to enhance \texttt{DeFog}.

\textbf{\textit{Application 1 - Deep learning based object classification using YOLOv3}}: 
The You Only Look Once (YOLO)\footnote{\url{https://pjreddie.com/darknet/yolo/}} is a real-time deep learning based object detection system that uses a single neural network for the entire image~\cite{yolo-01}. The image is decomposed into different regions and the bounding boxes and probabilities for each region are estimated. These bounding boxes are weighted by the predicted probabilities. The YOLOv3~\cite{yolov3-01} is a faster system compared to competing systems that implement Region-based Convolutional Neural Networks (R-CNN). The original application resizes a provided image asset and detects objects within the image using a pre-trained model. A labelled image is generated with percentage weights providing the accuracy of estimation. 

This system is an ideal candidate for the Fog. Fog applications that rely on this system may use the edge node to resize the original image (pre-process) so that the amount of data eventually transferred to the cloud beyond the edge is reduced. In addition, the objects can be detected at the edge to minimise communication latencies. The cloud server will need to send the pretrained model to the edge to allow the detection service to be hosted on the edge. 


\textbf{\textit{Application 2 - Speech-to-text conversion (PocketSphinx)}}:
PocketSphinx is an open-source large vocabulary, speaker-independent continuous speech recognition engine~\cite{pocketsphinx-01}. A supplied audio file (\texttt{.wav}) is converted to a defined language in text form using a pre-trained acoustic model to determine the source and destination language for speech-to-text conversion. The integrated assets are sourced from a large scale speech repository\footnote{\url{http://www.repository.voxforge1.org/downloads/SpeechCorpus}}. 
In the Fog application, the end user-device provides a \texttt{.wav} asset to the edge container. A pre-trained acoustic model is offloaded from the cloud to the edge to facilitate text-to-speech conversion. 

\textbf{\textit{Application 3 - Text-audio synchronisation or forced alignment (Aeneas)}}: 
The Aeneas tool automatically synchronises text and audio segments\footnote{\url{https://www.readbeyond.it/aeneas/docs/}}. Generating synchronisation maps for a list of text fragments and an audio file containing the relevant text is referred to as forced alignment. Aeneas determines the mapping between corresponding audio segment for each text asset supplied. 

In the Fog application, the end user-device transfers an audio file (\texttt{.wav}) to the edge node. Text segmentation occurs on the cloud and the text segment (\texttt{.xhtml}) is offloaded from the cloud to the edge where the forced alignment occurs. 


\textbf{\textit{Application 4 - Geo-location based online mobile game (iPokeMon)}}:
The original version of iPokeMon is a cloud-based online game that is similar to the popular virtual reality game, Pok\'emon Go\footnote{\url{http://www.pokemongo.com/}}. The user device interacts with the cloud server from the beginning to the completion of the game. The Fog version employed in \texttt{DeFog} is the use-case that is employed for validating research on resource management at the edge of the network~\cite{enorm-01}. 

In the Fog version, the user creation and verification requests from the user device are made to the cloud server, after which the cloud server manager makes a request for computing services on the edge node. If this request is accepted by the edge node, then the edge manager initialises a container for the iPokeMon edge server. The cloud manager deploys the iPokeMon edge server and clones (to the edge database) location specific data and user-specific data (of those who will be connected to the edge). User data rapidly changes when the game is played. For example, the GPS coordinates of the player and the Pok\'emons. The local view on the edge server is updated by frequent requests sent to the server. 

For experimentation, JMeter and Taurus are used to simulate players behaviour by generating synthetic workloads~\cite{enorm-01}. The user device supplies the \texttt{.jmx} file to JMeter. This allows for the automated generation of workloads for a range of concurrent users.

\textbf{\textit{Application 5 - Internet-of-Things (IoT) edge gateway application (FogLAMP)}}:
This is an open source IoT application that integrates cloud storage and sensors\footnote{\url{https://foglamp.readthedocs.io/en/latest/}}. FogLAMP interacts with endpoint sensors to collect and aggregate data. A text payload containing a curl command is transferred to the edge from endpoints. The service running on the edge executes the curl command by invoking a simulated API call that gathers the localised data. 

\textbf{\textit{Application 6 - Real-time face detection from video streams (RealFD)}}: 
The original application uses an end device with an embedded video camera to capture a continuous video stream. The stream is transmitted to a cloud server where faces are detected on individual video frames using Pillow\footnote{\url{https://pillow.readthedocs.io}} and OpenCV\footnote{\url{https://opencv.org}}. 


The server comprises the following three services for detecting faces from a frame of the video: (i) \textit{Grey-scale converter (GSC)} reduces the amount of computation done on an image that would otherwise be required if it was a colour image. The size of the stream is reduced by nearly a third, (ii) \textit{Motion Detector (MD)} is a data filtering service that performs a condition check on successive video frames to reduce computations on similar frames (for example a security feed in a static environment, such as a home or museum). (iii) \textit{Face Detector (FD)} is a computationally expensive service that identifies frontal faces in a video frame using machine learning.

The application can be distributed in the following four ways: (i) Cloud-only services -- all the services are deployed in the cloud; (ii) Fog-based pre-processing -- GSC is deployed at the edge and other services on the cloud; (iii) Fog-based data filtering -- GSC and MD are offloaded to the edge and FD is deployed on the cloud; (iv) Edge-only services -- all services are offloaded to the edge.

%% file: metrics.tex
Existing benchmarking methods that contain Fog conducive benchmarks either are workload specific benchmarking techniques, such as the speech-to-text benchmark\footnote{\url{https://github.com/Picovoice/stt-benchmark}} and TailBench~\cite{tailbench-01}, or are provider specific services benchmarking (AWS Greengrass and Microsoft Azure IoT Edge), such as EdgeBench\footnote{\url{https://github.com/akaanirban/edgebench}}~\cite{edgebench-01}. The motivation of \texttt{DeFog} is a general purpose Fog benchmarking method that can be used for a diverse range of workloads and at the same time provide both the target platform and application related benchmarks. 

Three types of metrics, namely platform, application specific and from external tools are considered. These provide a bird's eye view of the target platform, including network characteristics and applications running on the platform. 

\textbf{\textit{(i) Target Platform Metrics}}: \texttt{DeFog} gathers platform metrics for the three deployment modes (cloud-only, edge-only, cloud-edge). A short list of these is shown in Table~\ref{tab:platformmetrics}.

\begin{table}[ht]
  \begin{center}
    \caption{Platform metrics collected by \texttt{DeFog}}
    \label{tab:platformmetrics}
    \begin{tabular}{l|p{5.5cm}}
      \hline
      \textbf{Name} & \textbf{Description}\\
      \hline
      CPU Model Name & The model name of the target platform CPU\\
      No. of Cores & The total cores available on the CPU\\
      CPU Frequency & The frequency of the CPU\\
      System Uptime & The total time the system has been running\\
      Unzip time & The total time taken to unzip any assets (for example, a 34MB file for YOLOv3, which is the weights file)\\
      Download rate & The download rate (MB/sec) of assets (for example, 200MB model file for YOLOv3)\\
      System I/O & Speed of reading and writing in MB/sec\\
      \hline
    \end{tabular}
  \end{center}
\end{table}

\textbf{\textit{(ii) Fog Application Metrics}}:
\texttt{DeFog} provides insight into the six application benchmarks and the metrics are outlined in Table~\ref{tab:applicationmetrics}. These metrics provide insight into comparing the different deployment modes and are obtained in three categories: (1) Communication metrics, include the overheads of transferring assets and data payloads during the execution of the benchmark applications, (2) Computational performance metrics, include the time taken to execute a computational task, and (3) Concurrency metrics, which quantifies the impact of the workload on background noise and multiple workloads executing on the target platform.

\textbf{\textit{(iii) Metrics Gathered from External Tools}}: 
Two external tools, namely JMeter\footnote{\url{https://jmeter.apache.org/}} and Taurus\footnote{\url{https://gettaurus.org/}} are currently used by \texttt{DeFog}. The former is used to simulate multiple clients using a benchmark application, for example multiple users in the iPokeMon online game, The metrics obtained from JMeter is presented in Table~\ref{tab:jmetermetrics}. The end-to-end latency of the application when multiple users are running can be determined to identify whether the QoS of the application is met when the number of clients increases. 

\begin{table}[h]
  \begin{center}
    \caption{Metrics collected using JMeter in \texttt{DeFog}}
    \label{tab:jmetermetrics}
    \begin{tabular}{l|l}
      \hline
      \textbf{Name} & \textbf{Description}\\
      \hline
      User/ Thread & The total concurrent users/threads\\
      Latency & Response time latency for a specific endpoint\\
      \hline
    \end{tabular}
  \end{center}
\end{table}

Taurus gathers successful response count and average response time (shown in Table~\ref{tab:taurusmetrics}). By simulating synthetic users metrics such as standard deviation of response time can be generated, which offers insight into the effect of outliers on median response time. 

\begin{table}[h]
  \begin{center}
    \caption{Metrics collected using Taurus in \texttt{DeFog}}
    \label{tab:taurusmetrics}
    \begin{tabular}{p{3.3cm}|p{5cm}}
      \hline
      \textbf{Name} & \textbf{Description}\\
      \hline
      Concurrency & Average number of concurrent users\\
      Throughput & The total count of sample workloads\\
      Success/Fail & The total count of successful workloads\\
      Average Response Time (RT) & Average response time of service\\
      Standard Deviation of RT & Standard deviation of the response time\\
      Average Latency & Average latency time for the return trip\\
      \hline
    \end{tabular}
  \end{center}
\end{table}

\begin{table*}[ht]
  \begin{center}
    \caption{Fog application metrics collected by \texttt{DeFog}}
    \label{tab:applicationmetrics}
    \begin{tabular}{l|p{12.5cm}}
      \hline
      \textbf{Name} & \textbf{Description}\\
      \hline
      Execution Time &  Time taken to complete a computational task ($ET$)\\
      Time in Flight & Time taken to transfer a workload to the cloud or edge ($T_1$)\\
      S3 Transfer Time & Time taken to upload results data to the S3 bucket ($T_2$)\\
      Results Transfer Time & Time taken to return results data to the observing system ($T_3$)\\
      Computation Latency & Total time taken for running the benchmark, including execution time ($RTT = T_1 + ET + T_3$)\\
      Computation Cost & Estimated cost of the computational task ($Cost$, using the AWS pricing strategy)\\
      Real Time Factor & Rate of speech recognition ($RTF = ET / file\_length$)\\
      Bytes Up & Total bytes transferred to the cloud or edge\\ 
      Bytes Down & Total bytes transferred from the cloud or edge\\
      Bytes Down (cloud-edge) & Total bytes transferred from the cloud to the edge\\
      Bytes Up per sec & Upload rate in bytes per second\\
      Bytes Down per sec & Download rate in bytes per second\\ 
      Bytes Down (cloud-edge) per sec  & Download rate in bytes per second (cloud or edge) download rate\\
      Cloud-edge Transfer Time & Time taken to transfer assets from the cloud to the edge ($T_4$)\\ 
      Communication Latency & Total time taken for assets to be transferred throughout the benchmarking process ($CL = T_1 + T_3$)\\
      Complete Computation Latency & Return trip time including the time taken to transfer assets from the cloud to the edge\\
      Complete Communication Latency & Total time taken for the transfer of all payloads including the cloud asset\\ 
      \hline
    \end{tabular}
  \end{center}
\end{table*}

%% file: experiments.tex
The setup for capturing metrics for the three deployment modes and the results obtained are presented in this section. Only a subset of metrics and results obtained are presented, given that \texttt{DeFog} generates an exhaustive list of metrics and results. Moreover, aggregate metrics, which is a sum of a collection of individual metrics is considered. For example, computation latency is the sum of the time taken to transfer a service to the cloud or edge, time taken to complete executing a service, and the time taken to return output data to the observing system. It would not be possible to present and discuss the individual metrics within the scope of this paper.

\subsection{Implementation and Setup}
\label{sec:implementation}
\input{implementation.tex}

\subsection{Results}
\label{sec:results}
\input{results}

%% file: implementation.tex

The cloud resource used is an AWS EC2 instance that is set up in the Dublin \texttt{eu-west-1} region. 
The AWS command line interface (CLI) is used to set up Information Access Management (IAM) credentials. When an IAM user is instantiated, public and private keys are configured to allow for secure remote access and control of the EC2 instance.
Secure Shell (SSH) is used to ensure a secure connection is maintained with the cloud.
A data volume is allocated to the instance on S3 for adequate space to build several distinct applications simultaneously. The latest Docker dependencies are installed to allow for seamless deployment and execution. 

\begin{figure}[]
\begin{center}
	\subfloat[Odroid XU4]
	{\label{fig:edgeresource1a}
	\includegraphics[width=0.225\textwidth]
	{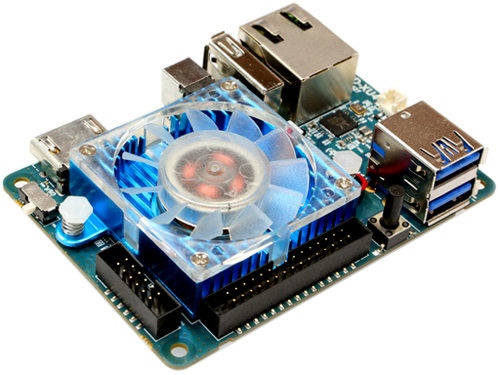}}
\hfill
	\subfloat[Raspberry Pi 3 Model B]
	{\label{fig:edgeresource1b}
	\includegraphics[width=0.225\textwidth]
	{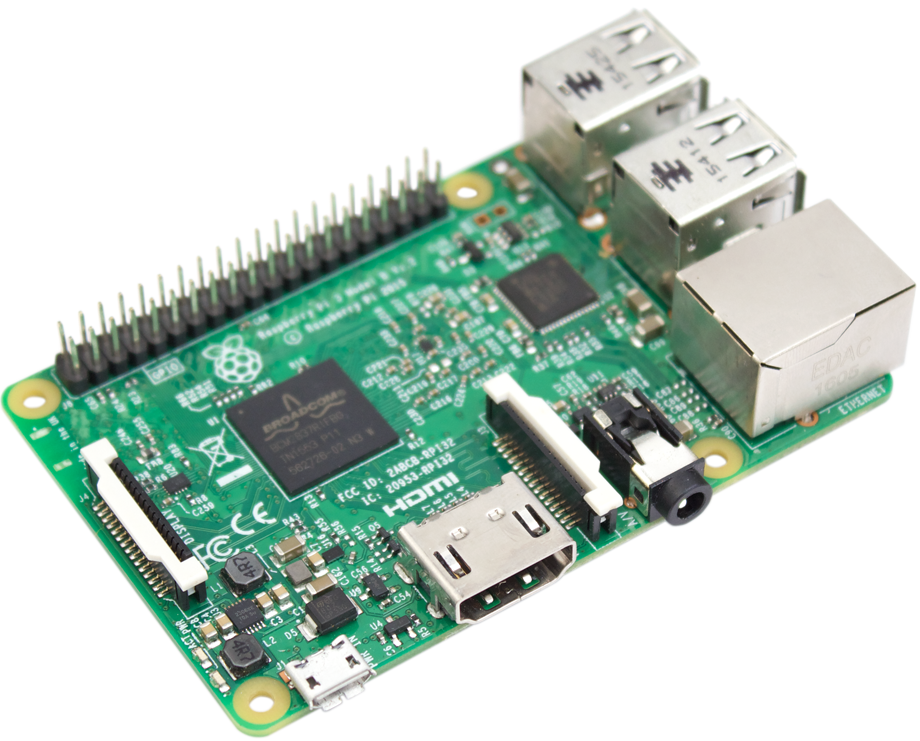}}
\end{center}
\caption{Single board computers used as edge resources.}
\label{fig:edgeresources}
\end{figure}

Given that edge resources are hardware limited when compared to the cloud, this research employs embedded single board computers, namely Odroid XU4 and a Raspberry Pi 3 Model B as shown in Figure~\ref{fig:edgeresources}. These single board computers have resources comparable to the compute that is available on a base station~\cite{enorm-01}. 
The Odroid board has 2~GB of DRAM memory, and one ARM Big.LITTLE architecture Exynos 5 Octa processor running Ubuntu 14.04 LTS.
The Raspberry Pi has 
1~GB of RAM memory, and a Quad Core 1.2GHz Broadcom BCM2837 64 bit CPU running Raspbian. 

The applications are deployed using the container technology, specifically Docker\footnote{\url{https://www.docker.com/}}. Docker 17.12.1-ce is used to automate building and deploying the application. Application specific Dockerfiles install the relevant dependencies and packages when the containers are build. Container instances are then run using the same application image. A combination of Python and bash scripting is used along with Dockerfiles to execute the benchmarks.

For each benchmark a data payload is transferred from the observing system to the edge or cloud. Fog deployment requires assets to be offloaded from the cloud to edge. Data generated on the cloud and/or edge is transferred to the S3 bucket and observing system. 

YOLO, PocketSphinx, iPokeMon, Aeneas and RealFD are benchmarked across all three deployment modes. Taurus and Apache JMeter are used to simulate concurrent users. FogLAMP does not require a cloud asset, and is not used in the Fog mode, but for the cloud-only and edge-only deployment modes. 

The aim of the experiments are to demonstrate the metrics that can be obtained to compare the relative computational and communication performance when the edge is leveraged. The experiments are performed on the three deployment modes (cloud-only, edge-only, and cloud-edge) for gaining insight to \textit{\textbf{Q1}} that was posed in Section~\ref{sec:introduction}. The results obtained from the RealFD application (multiple services that can be moved to the edge) highlight the use of \texttt{DeFog} to answer \textit{\textbf{Q2}}. The experiments are carried out on two edge resources (Odroid XU4 and Raspberry Pi 3), which enhances our knowledge of the performance of the benchmark in relation to \textit{\textbf{Q3}}.  
The experiments are carried out for single and multiple users as well as when the edge resource is stressed to reflect real-world deployments. Each experiment was executed 25 times.

%% file: results.tex
\begin{figure*}[ht]
\begin{center}
	\subfloat[Communication latency]
	{\label{fig:communicationlatency}
	\includegraphics[width=0.495\textwidth]
	{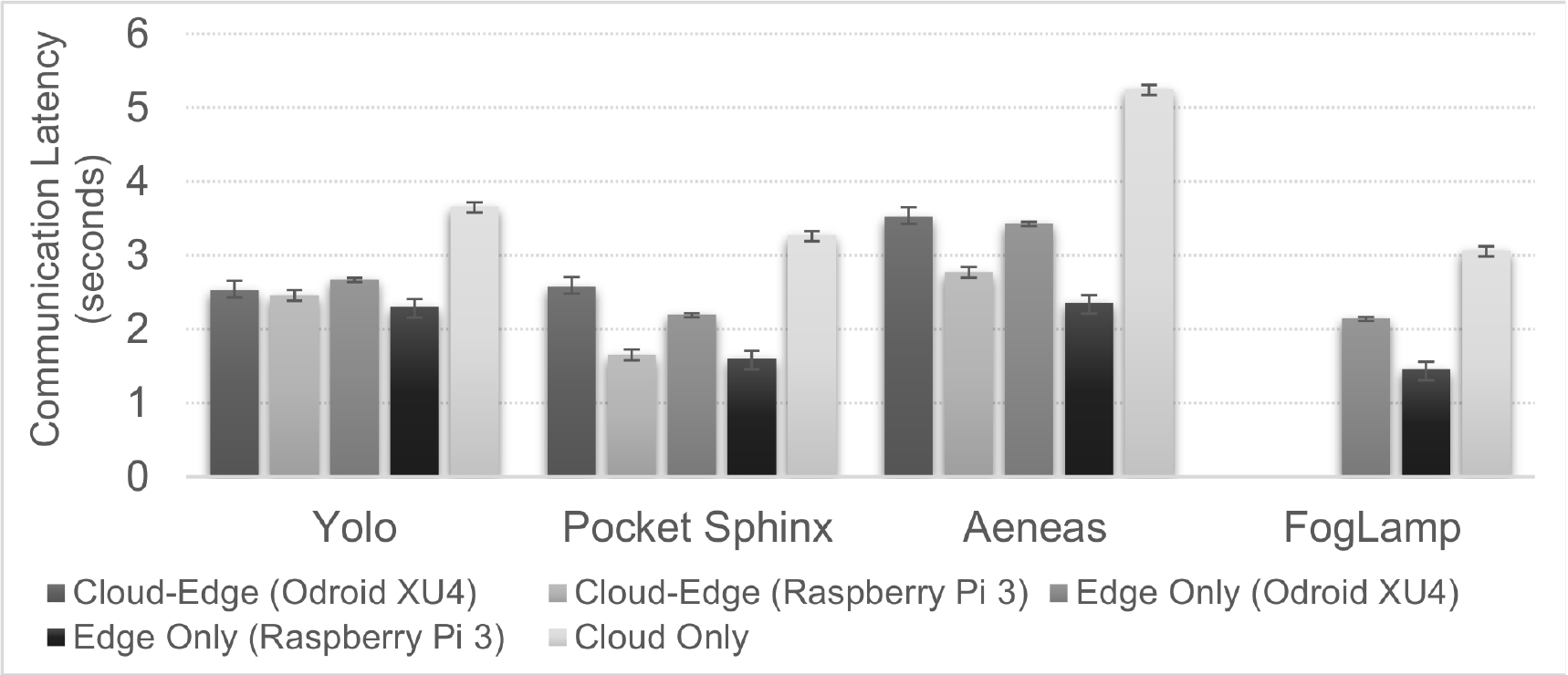}}
\hfill
	\subfloat[Computation latency]
	{\label{fig:computationlatency}
	\includegraphics[width=0.495\textwidth]
	{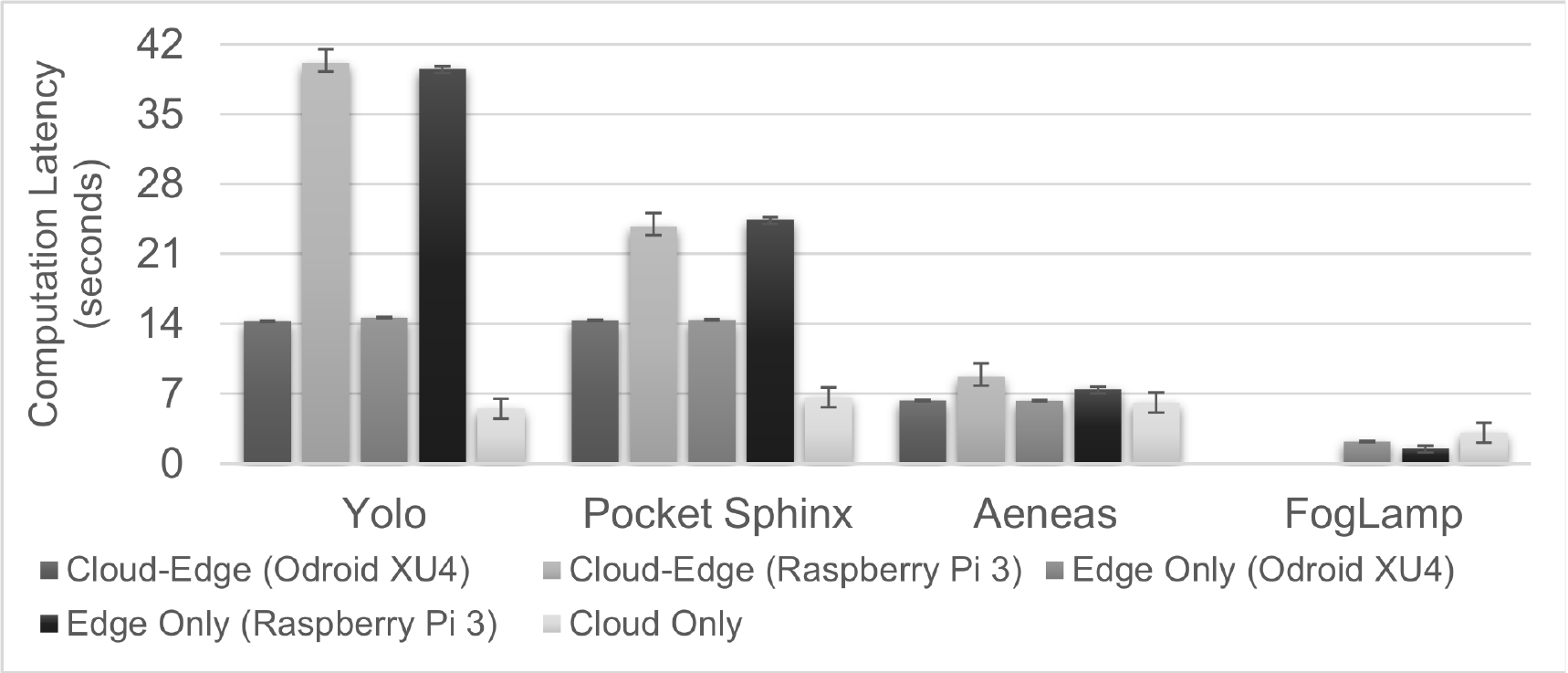}}
\end{center}
\caption{Latencies of applications for different deployment modes.}
\label{fig:latencies}
\end{figure*}

\begin{figure}
	\centering
	\includegraphics[width=0.495\textwidth]{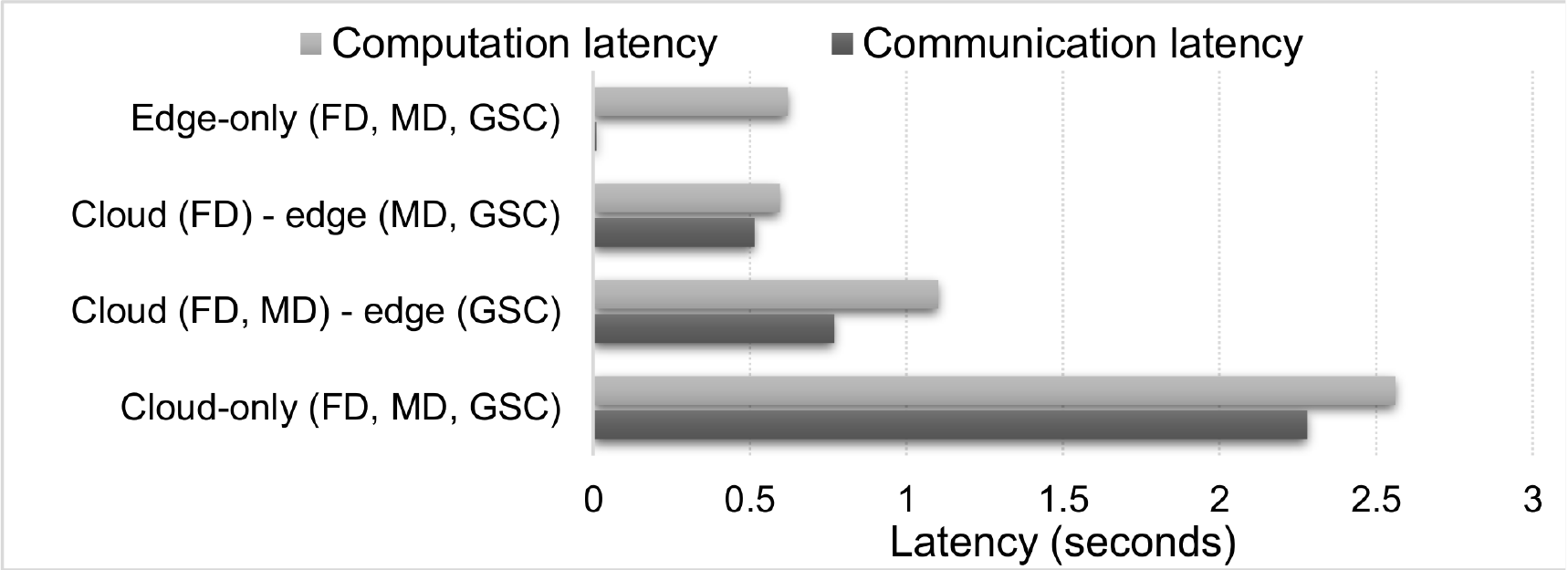}
	\caption{Latencies of RealFD for different combination of services across the cloud and the edge (using Odroid XU4).}
\label{fig:realfdlatencies}
\end{figure}

\begin{figure}[ht]
\begin{center}
	\subfloat[On Odroid XU4]
	{\label{fig:stresscommunicationodroid}
	\includegraphics[width=0.233\textwidth]
	{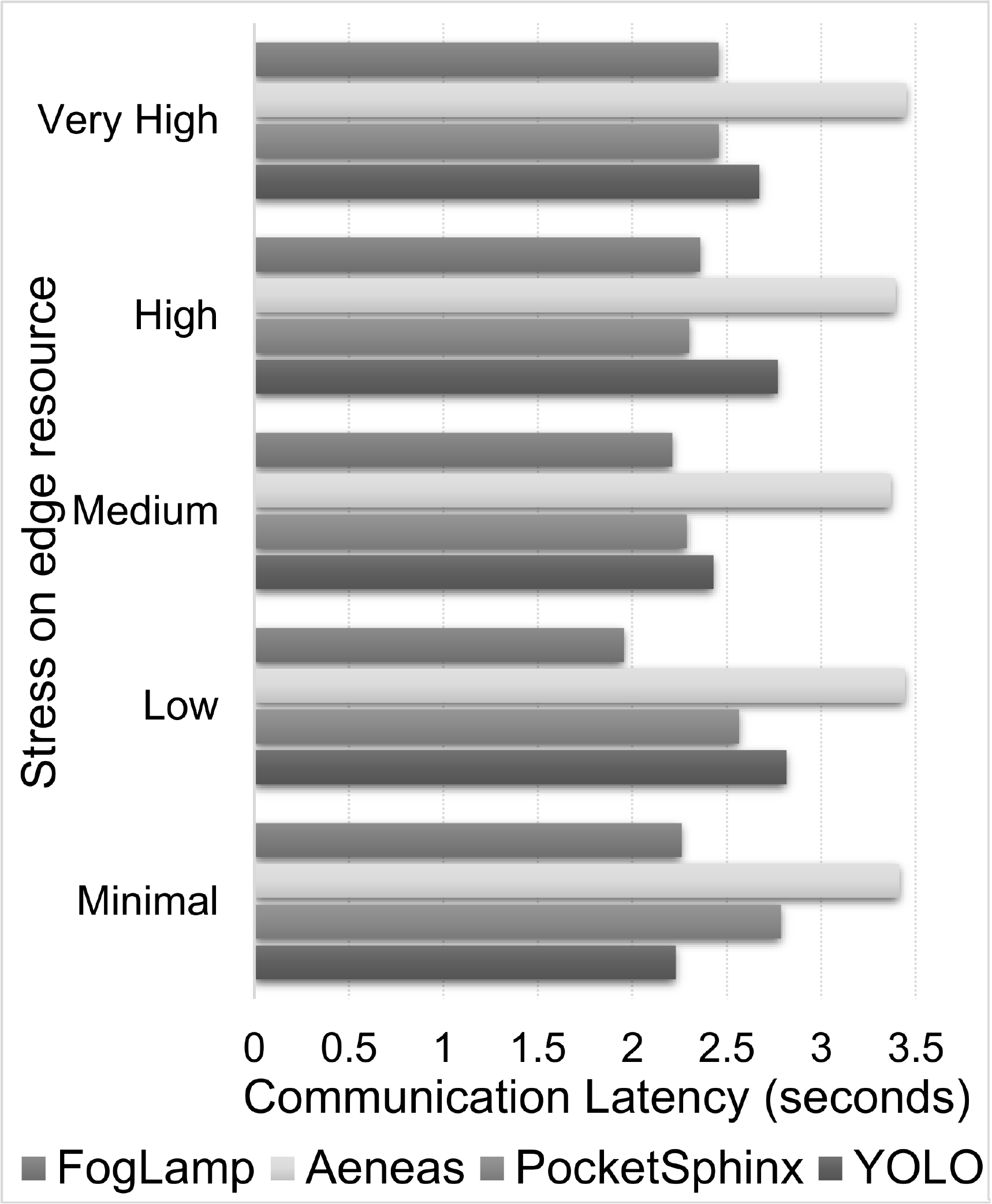}}
\hfill
	\subfloat[On Raspberry Pi 3]
	{\label{fig:stresscommunicationraspberry}
	\includegraphics[width=0.233\textwidth]
	{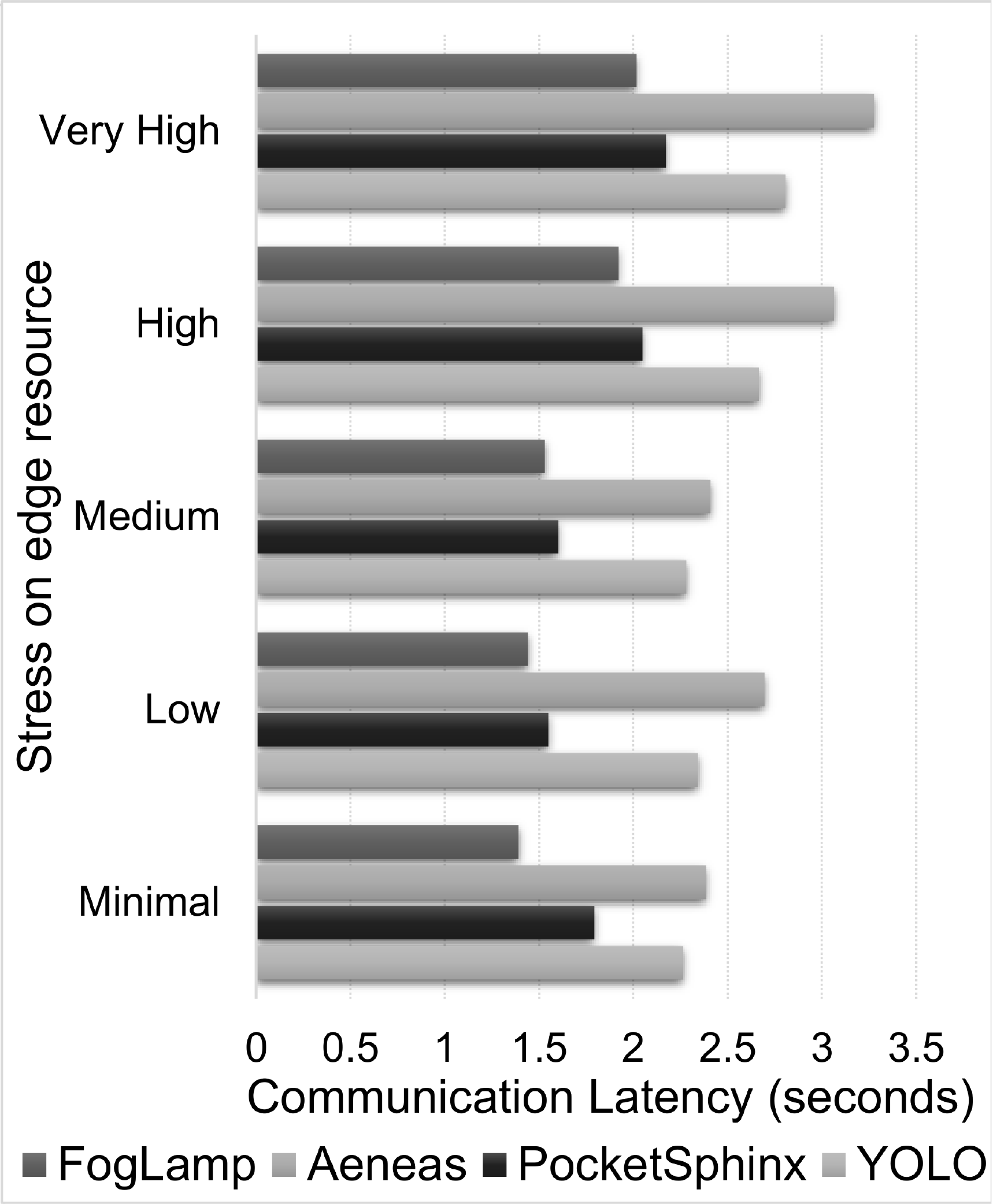}}
\end{center}
\caption{Communication latency of benchmark applications when the edge resource is stressed.}
\label{fig:stresscommunicationlatencies}
\end{figure}

\begin{figure}[]
\begin{center}
	\subfloat[On Odroid XU4]
	{\label{fig:stresscomputationodroid}
	\includegraphics[width=0.233\textwidth]
	{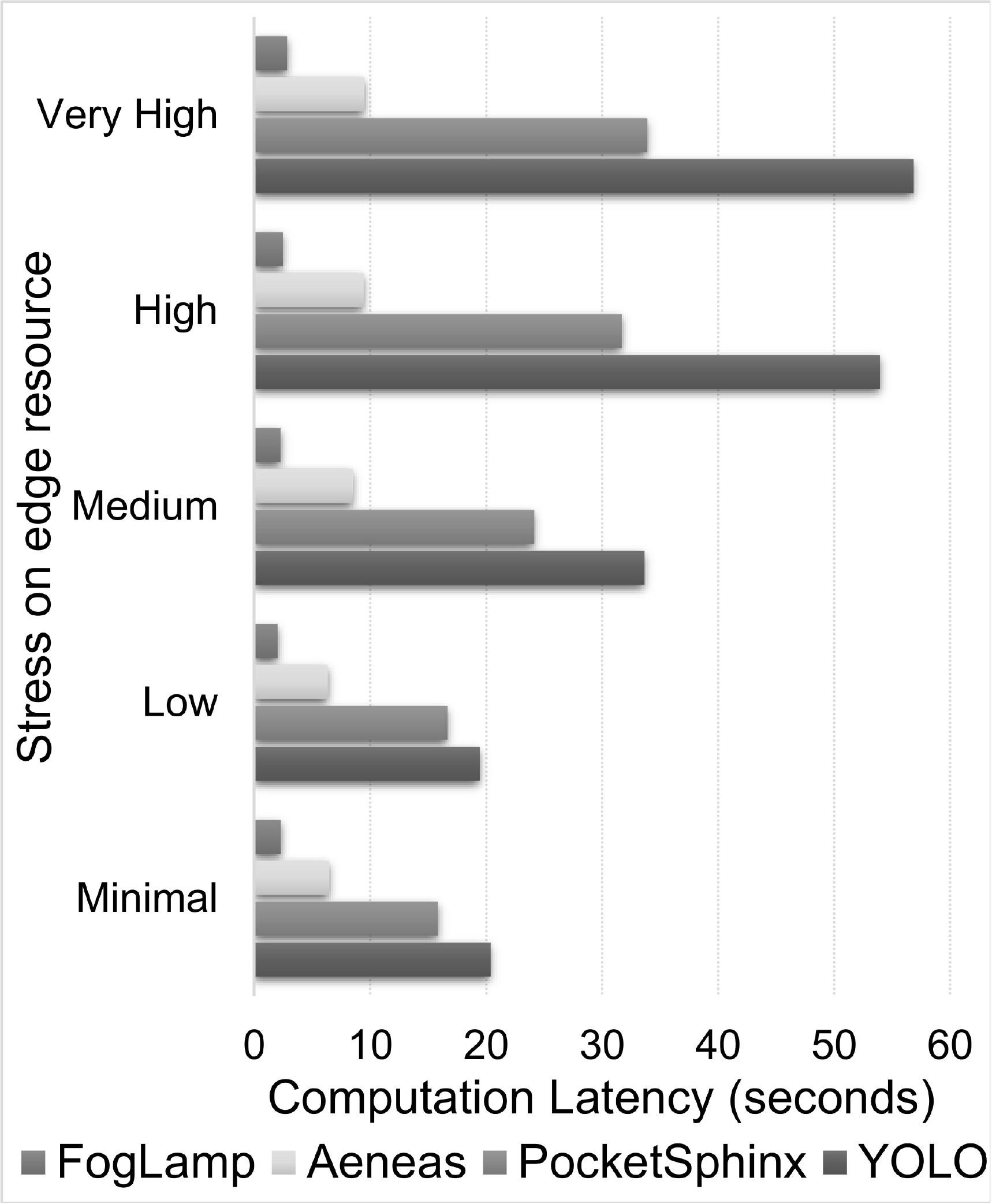}}
\hfill
	\subfloat[On Raspberry Pi 3]
	{\label{fig:stresscomputationraspberry}
	\includegraphics[width=0.233\textwidth]
	{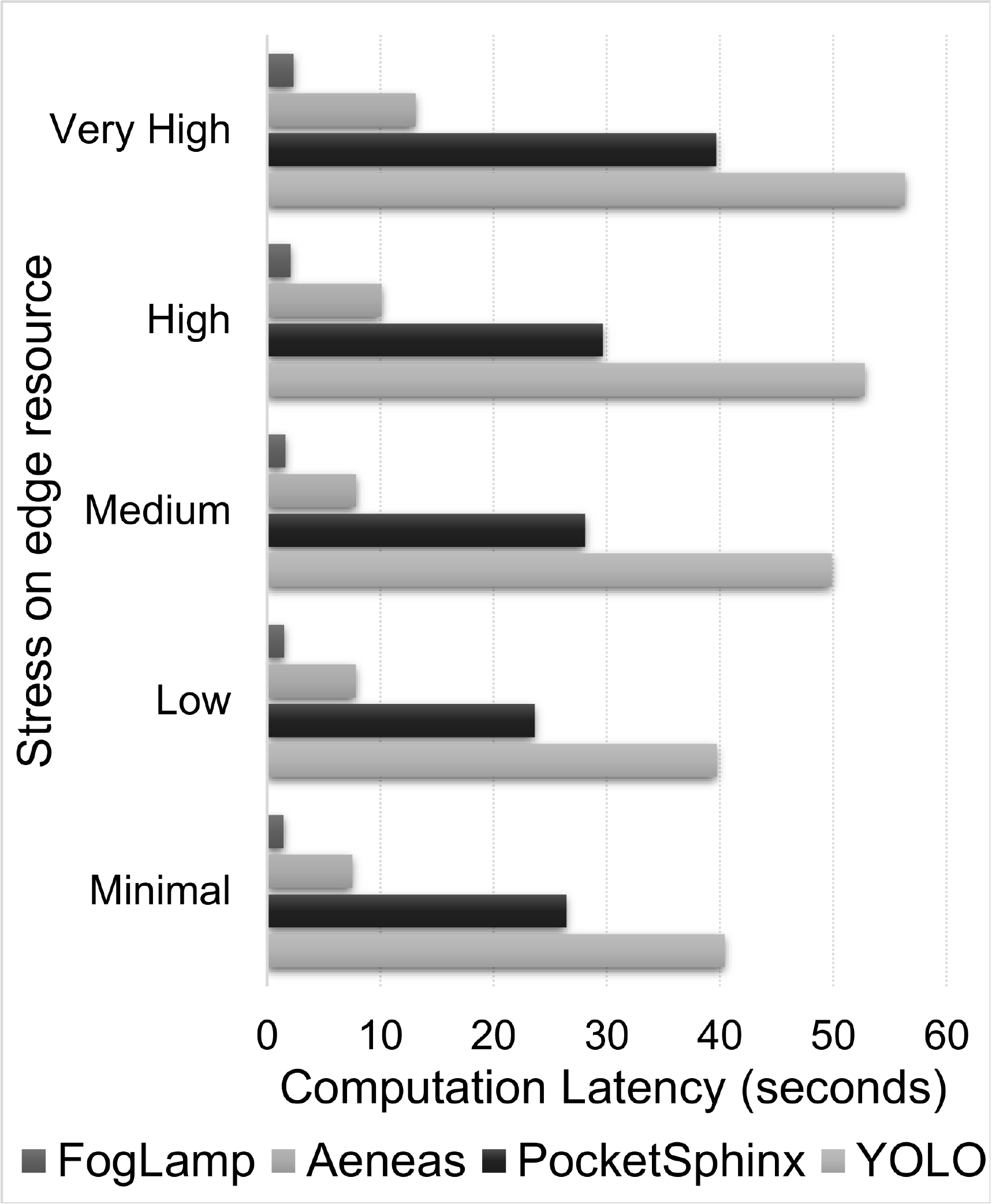}}
\end{center}
\caption{Computation latency of benchmark applications when the edge resource is stressed.}
\label{fig:stresscomputationlatencies}
\end{figure}

\begin{figure}[]
\begin{center}
	\subfloat[Communication latency]
	{\label{fig:stressrealfdcommunication}
	\includegraphics[width=0.495\textwidth]
	{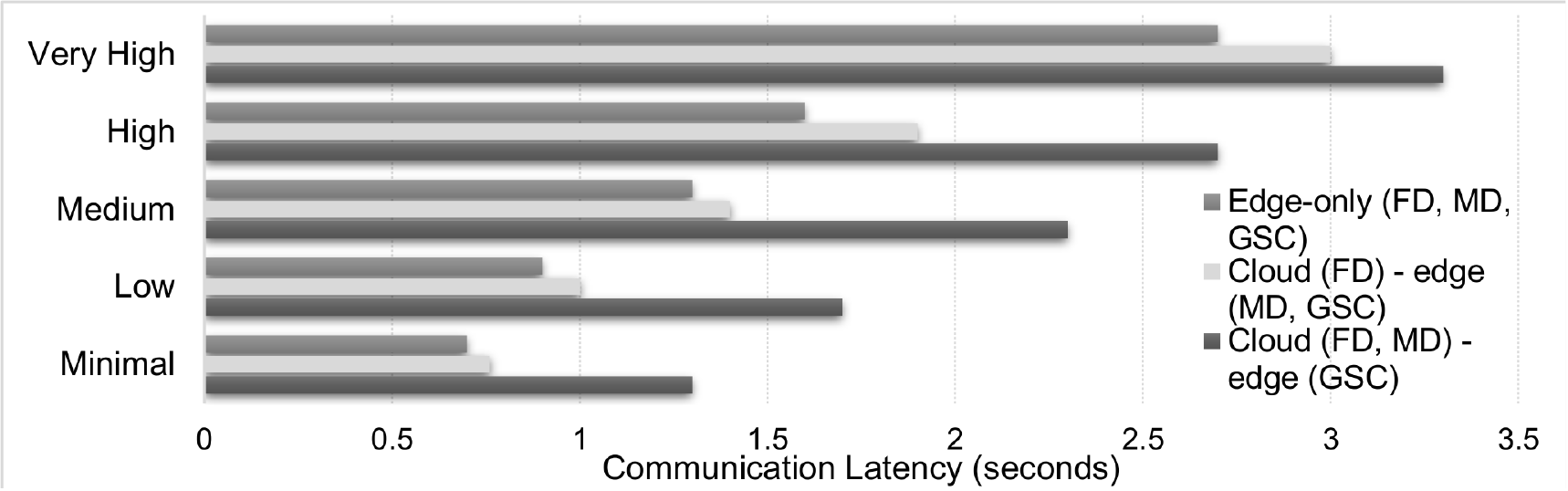}}

	\subfloat[Computation latency]
	{\label{fig:stressrealfdcomputation}
	\includegraphics[width=0.495\textwidth]
	{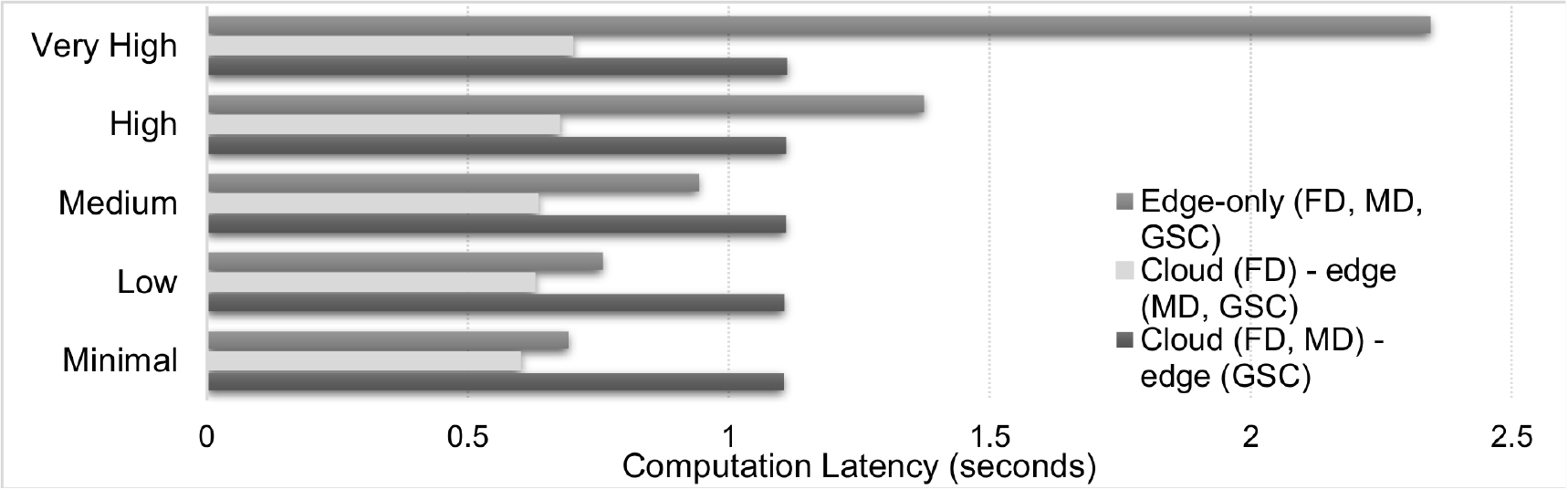}}
\end{center}
\caption{Latencies of the RealFD benchmark when the edge resource is stressed.}
\label{fig:stressrealfd}
\end{figure}

\begin{figure}
	\centering
	\includegraphics[width=0.47\textwidth]{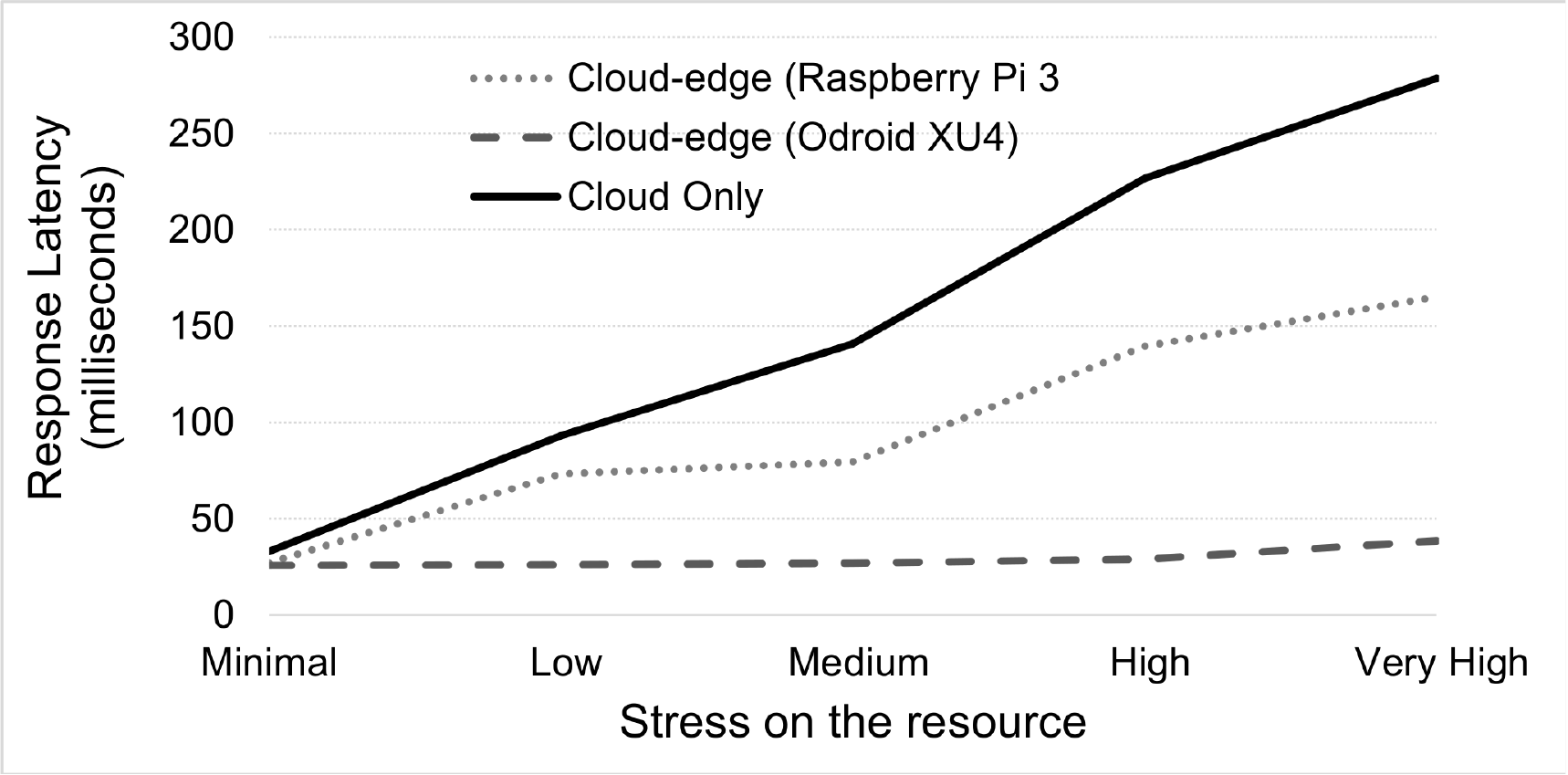}
	\caption{Impact of edge resource stress on response latency for iPokeMon.}
\label{fig:jmeterstress}
\end{figure}

The results obtained are presented as: (i) application latencies for different deployments, (ii) impact of stressing the edge on latencies, and (iii) impact of concurrent users on latencies. 

\subsubsection{Application latencies for different deployment modes}
Figure~\ref{fig:latencies} show the communication and computation round trip times for the three deployment modes, namely the cloud-only, edge-only, and cloud-edge (Fog) deployments. The standard deviation of the executions is highlighted in the observed results. In these executions, the edge resources are exclusively used by the benchmark application. The communication latency (Figure~\ref{fig:communicationlatency}) is consistently lower for all applications running on the edge when compared to the cloud. The computation latency (Figure~\ref{fig:computationlatency}) is significantly larger for applications, such as YOLO and PocketSphinx, on the edge compared to the cloud. This is because the edge resources employed in this research are hardware limited compared to a cloud resource. The computational cost of these applications exceeds the gains in communication latencies on the edge. There is a slight increase in the latency times for the cloud-edge deployment modes when compared to the edge only deployment, which is due to the time needed to transfer assets from the cloud to the edge. Aeneas and FogLAMP show comparable and sometimes lower computational latency on the edge when compared to the cloud. The lack of performance gain for the former two applications may be due to the specific manner in which the application is partitioned. 

Figure~\ref{fig:realfdlatencies} shows the computation and communication latencies for different combination of services of the RealFD application (FD, MD, GSC) on the three deployment modes. For this application, the results obtained using the Odroid XU4 are presented. Communication latency in this figure is the single trip time taken (not round trip) and the computation latency is the time taken to process a single video frame. There are two potential deployment options across the cloud and edge - FD on the cloud, and MD and GSC on the edge, or alternatively FD and MD on the cloud, and GSC on the edge. There is a difference in the performance of the two Fog deployments and \texttt{DeFog} highlights this variation for the benchmark. 


\subsubsection{Impact of stressing edge resources on application latencies}
For this experiment Gaussian workloads are simulated on the edge nodes in the cloud-edge deployment mode for YOLO, PocketSphinx, Aeneas, and RealFD and the edge-only deployment mode for FogLAMP. The motivation is to simulate a real world multi-tenant distributed system where there are competing workloads residing on the same edge under variable network conditions. The \texttt{stress}\footnote{\url{https://people.seas.harvard.edu/~apw/stress/}} package is used to stress the edge resource by simulating computations in the background. The \texttt{stress-ng}\footnote{\url{https://kernel.ubuntu.com/git/cking/stress-ng.git/}} package is used to stress the network. These packages use stressors to subject the computing cores and network to various levels of stress. 

In this paper, we explicitly define minimal stress when one CPU core of the edge resource is stressed. The network is stressed by transferring a file of size 256MB at roughly 21740 bytes per second. For \textit{low stress}, two CPU cores are stressed. For \textit{medium} and \textit{high stress}, three CPU cores and four CPU cores are stressed respectively. For \textit{very high stress} all CPU cores are stressed and the RAM memory is stressed using two stressor processes.  

The communication latency of benchmark applications when network bandwidth is stressed is shown in Figure~\ref{fig:stresscommunicationlatencies} (for the RealFD application on Odroid XU4 is shown in Figure~\ref{fig:stressrealfdcommunication}). Applications that transfer larger amounts of data, such as Aeneas, are affected. Figure~\ref{fig:stresscomputationlatencies} shows the trend in the computation latencies of the applications when the computing cores are stressed (Figure~\ref{fig:stressrealfdcomputation}). As expected there is a significant increase in the computation latencies. Computationally less intensive applications, such as Aeneas and FogLAMP have less effect with stressed CPU cores. It is immediately inferred that applications demonstrate different trends when edge resources are stressed. For RealFD it is noted that different combination of services across the cloud and the edge have different performance. 
For the iPokeMon application as shown in Figure~\ref{fig:jmeterstress}, the edge node is subject to similar stress as above. There is an improvement in the response latency of the cloud-edge deployment by over 7 times when compared to the cloud-only deployment when the stress as defined in this paper is very high.


\begin{figure*}[ht]
\begin{center}
	\subfloat[Communication latency on Odroid XU4]
	{\label{fig:userscommunicationodroid}
	\includegraphics[width=0.236\textwidth]
	{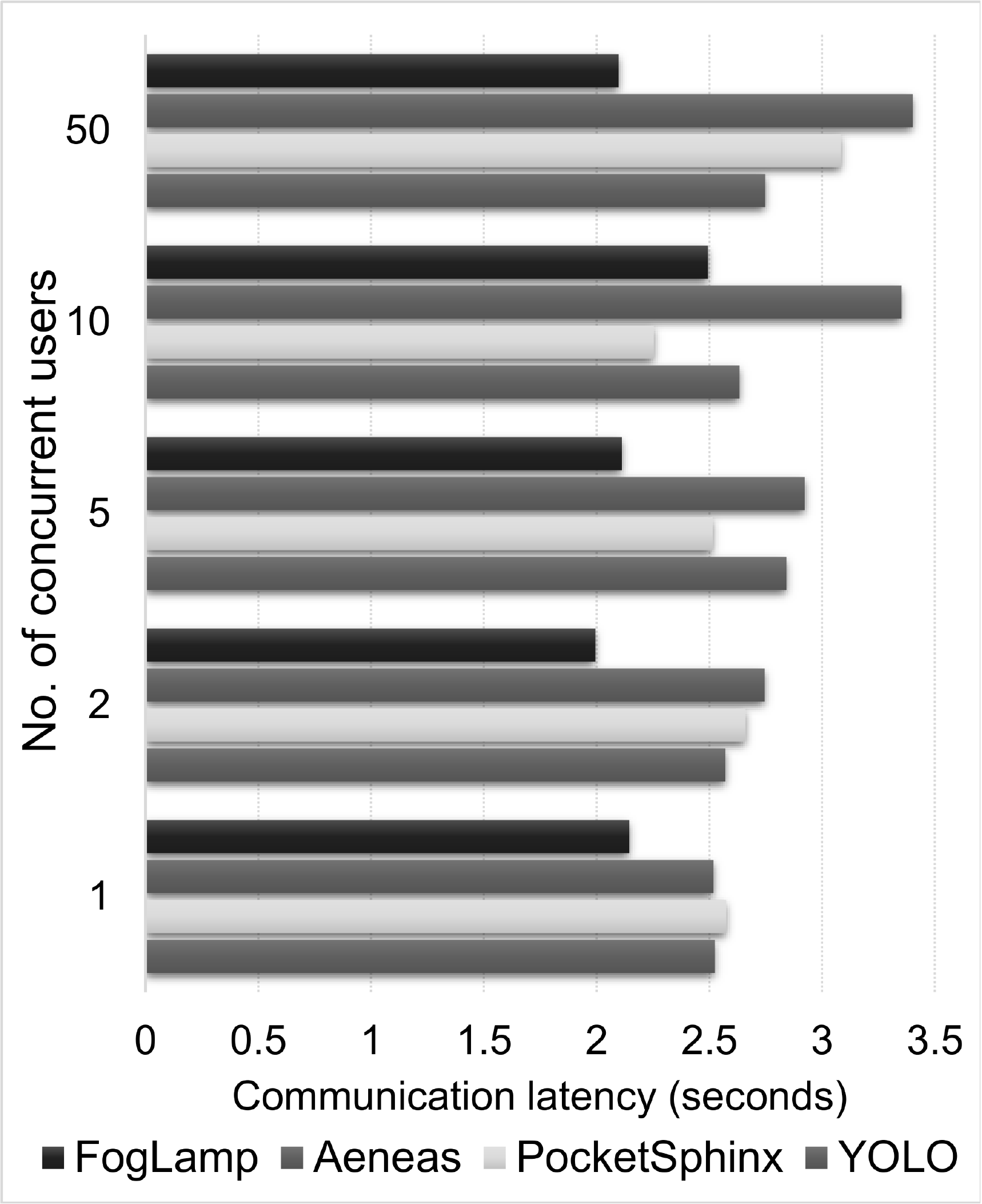}}
\hfill
	\subfloat[Communication latency on Raspberry Pi 3]
	{\label{fig:userscommunicationraspberry}
	\includegraphics[width=0.236\textwidth]
	{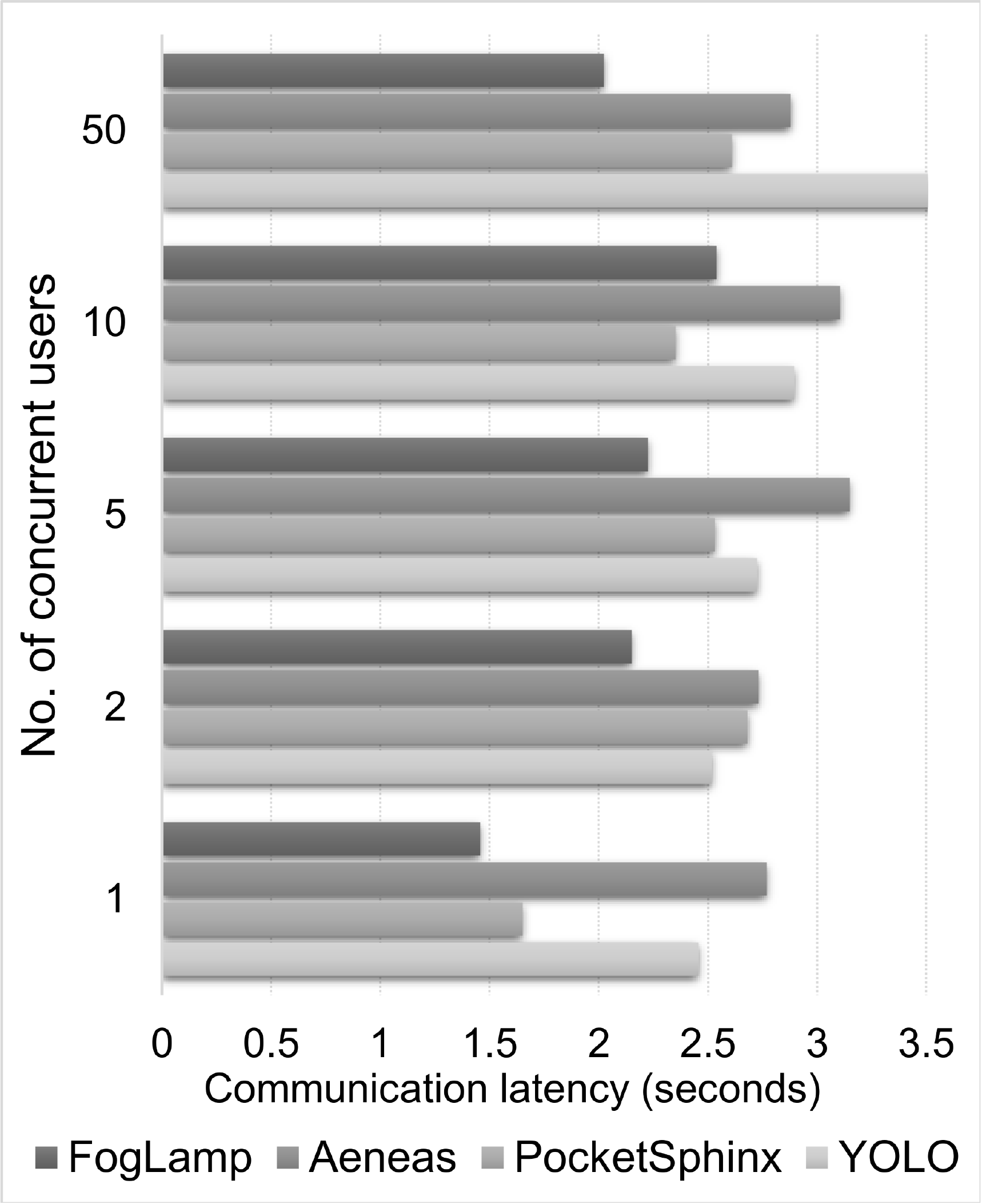}}
\hfill
	\subfloat[Computation latency on Odroid XU4]
	{\label{fig:userscomputationodroid}
	\includegraphics[width=0.236\textwidth]
	{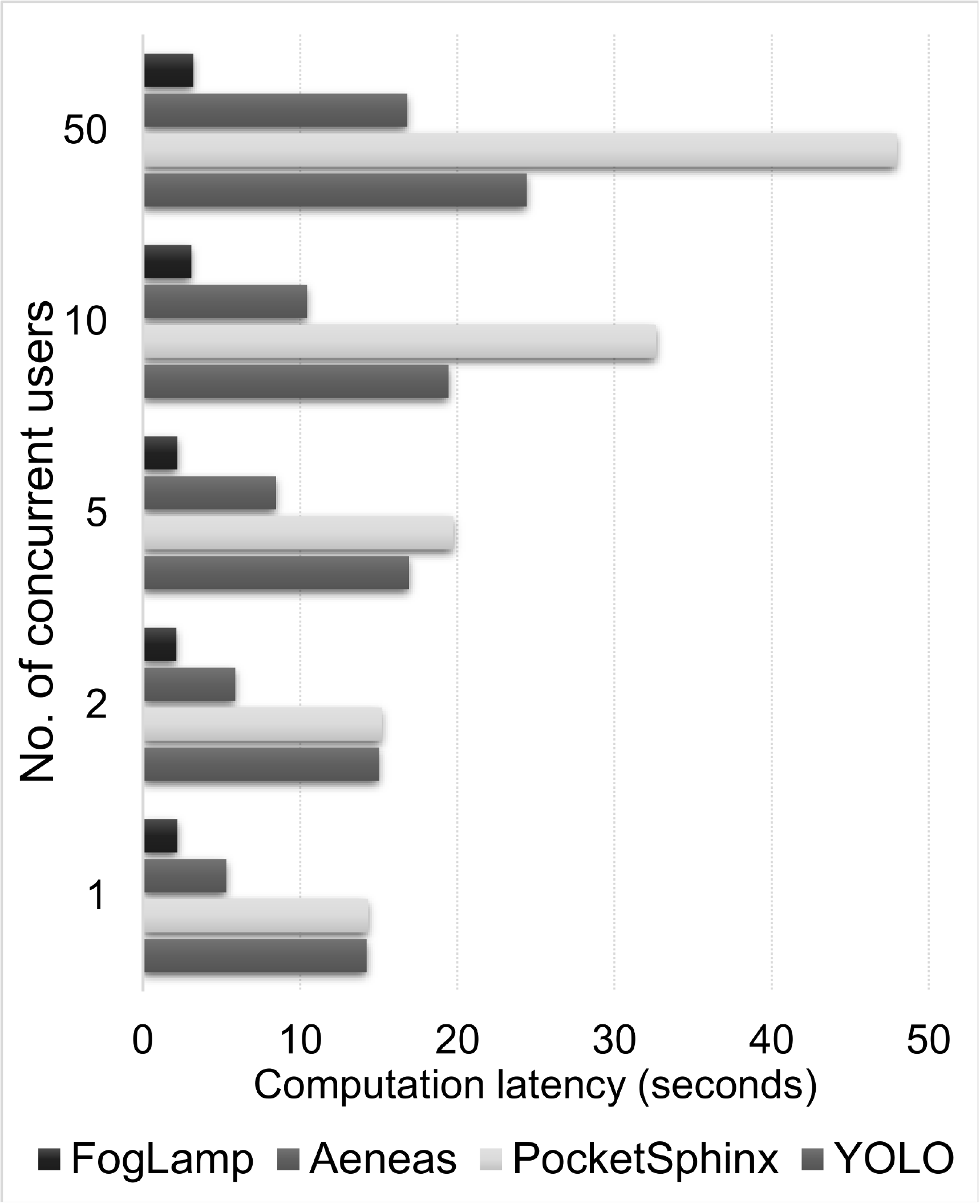}}
\hfill
	\subfloat[Computation latency on Raspberry Pi 3]
	{\label{fig:userscomputationraspberry}
	\includegraphics[width=0.236\textwidth]
	{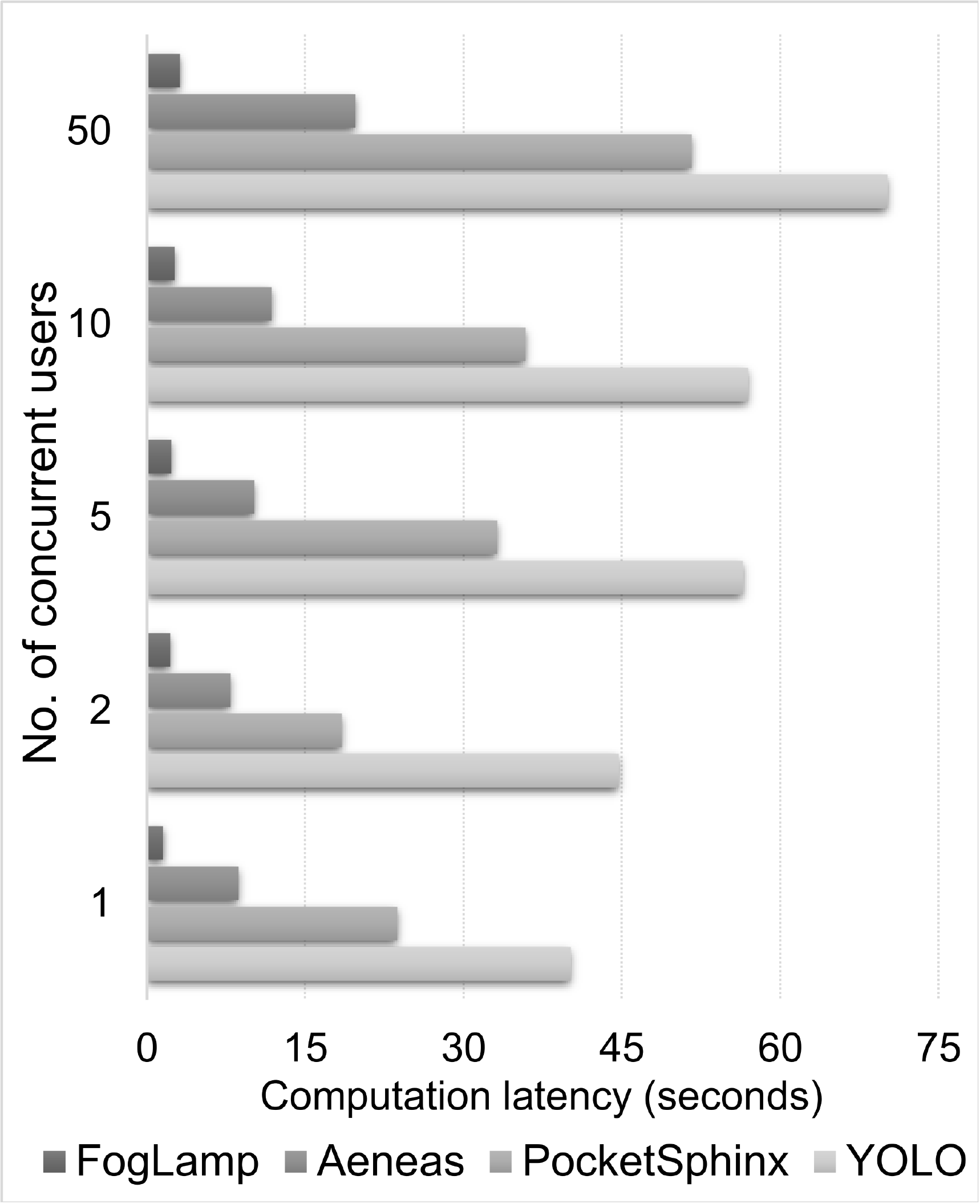}}
\end{center}
\caption{Impact of concurrent users on latency of benchmark applications.}
\label{fig:userslatencies}
\end{figure*}

\begin{figure*}[ht]
\begin{center}
	\subfloat[On Odroid XU4]
	{\label{fig:usersbytestransferodroid}
	\includegraphics[width=0.48\textwidth]
	{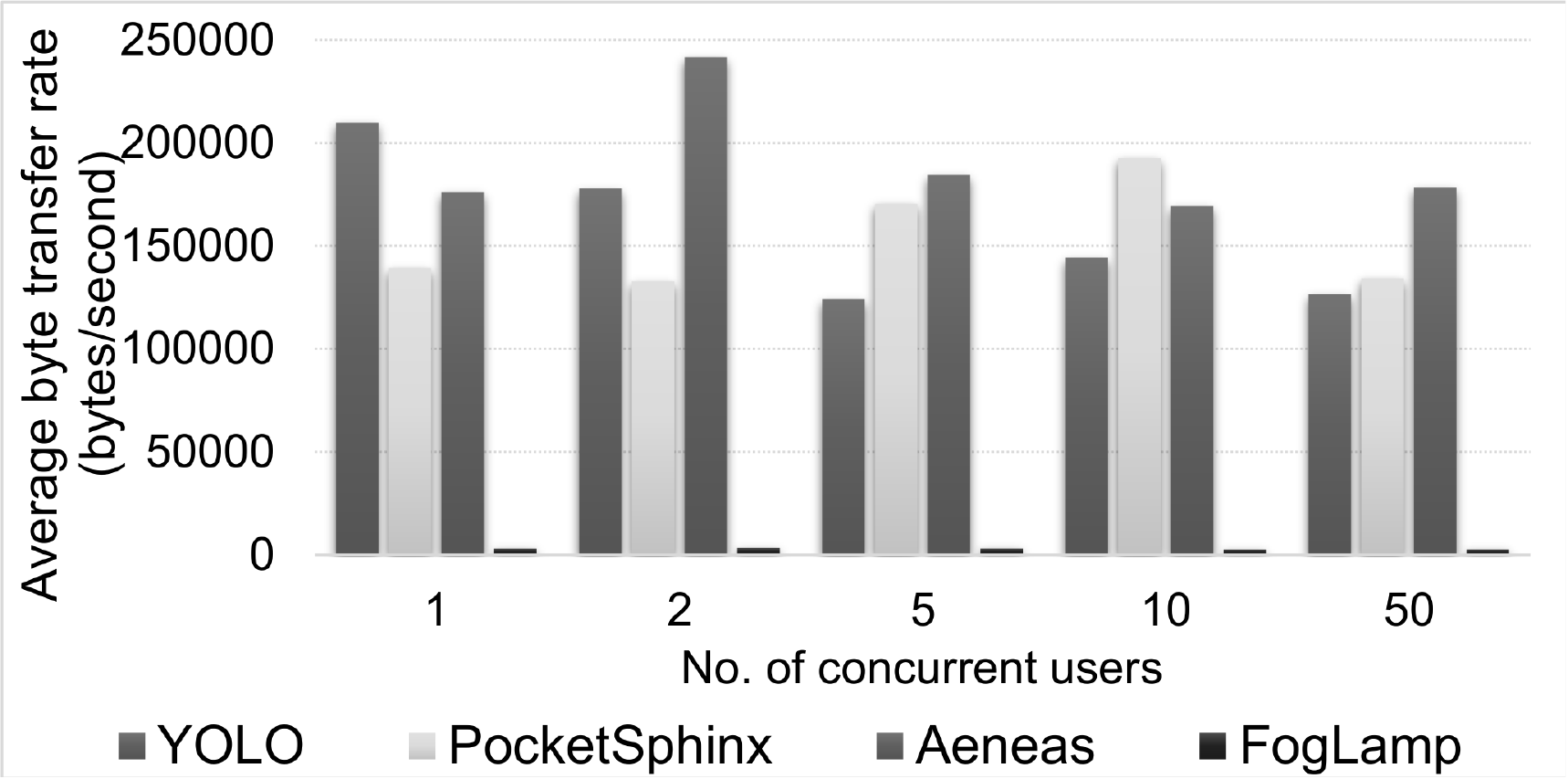}}
\hfill
	\subfloat[On Raspberry Pi 3]
	{\label{fig:usersbytestransferraspberry}
	\includegraphics[width=0.48\textwidth]
	{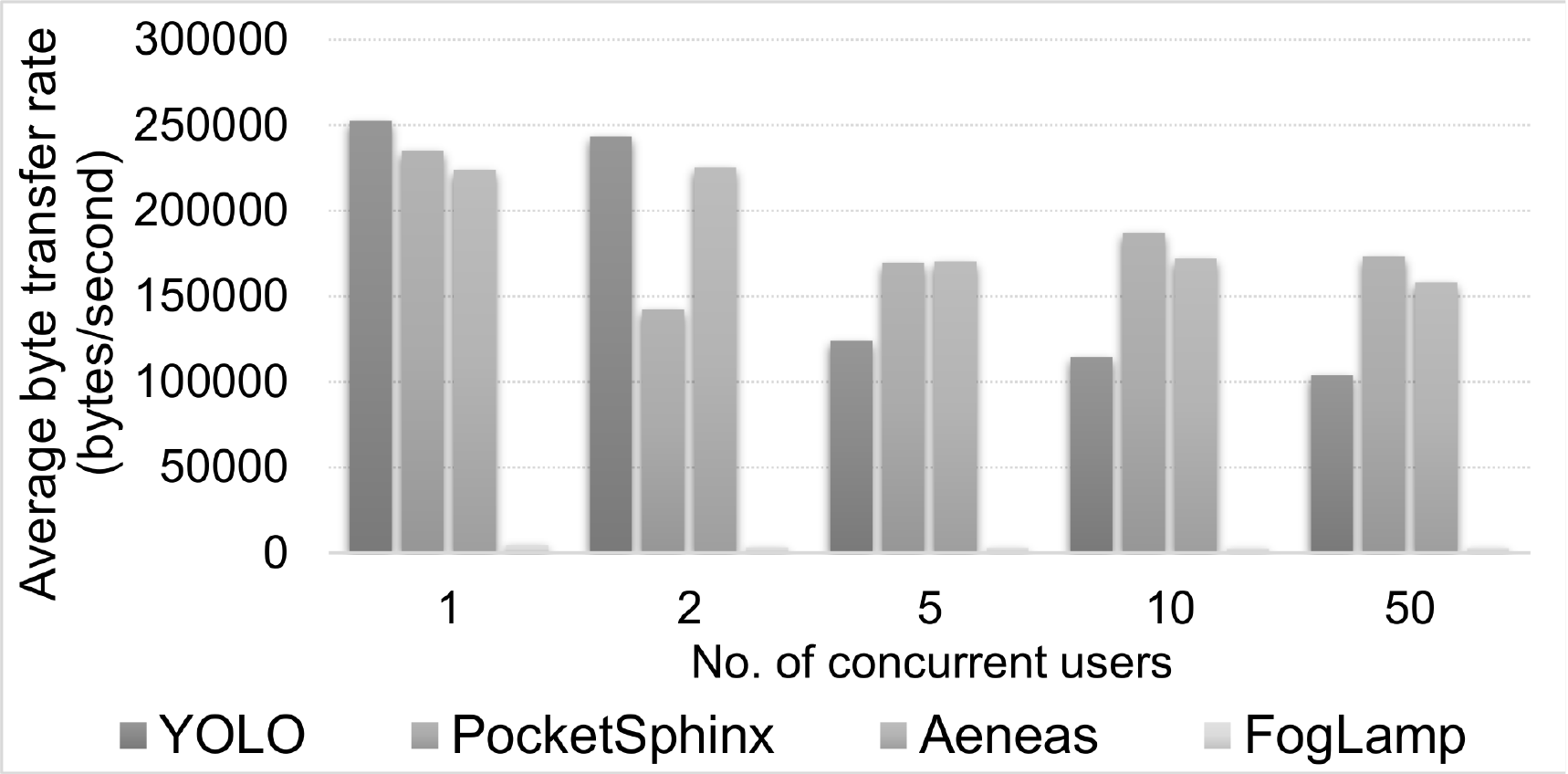}}
\end{center}
\caption{Impact of concurrent users on average bytes transferred.}
\label{fig:usersbytestransfer}
\end{figure*}

\begin{figure}
	\centering
	\includegraphics[width=0.48\textwidth]{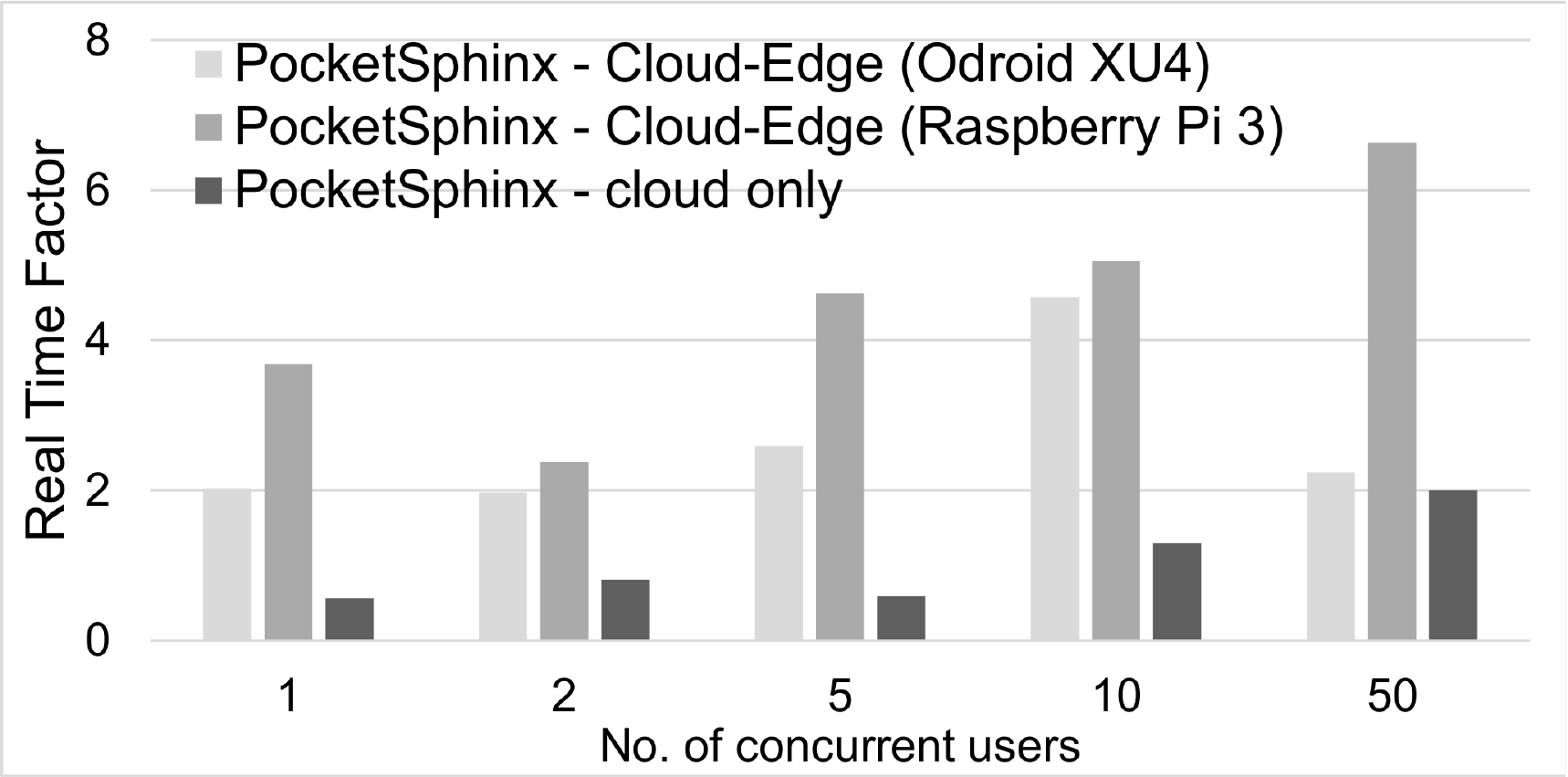}
	\caption{Impact of concurrent users on real time factor for PocketSphinx.}
\label{fig:usersrtf}
\end{figure}

\begin{figure}
	\centering
	\includegraphics[width=0.48\textwidth]{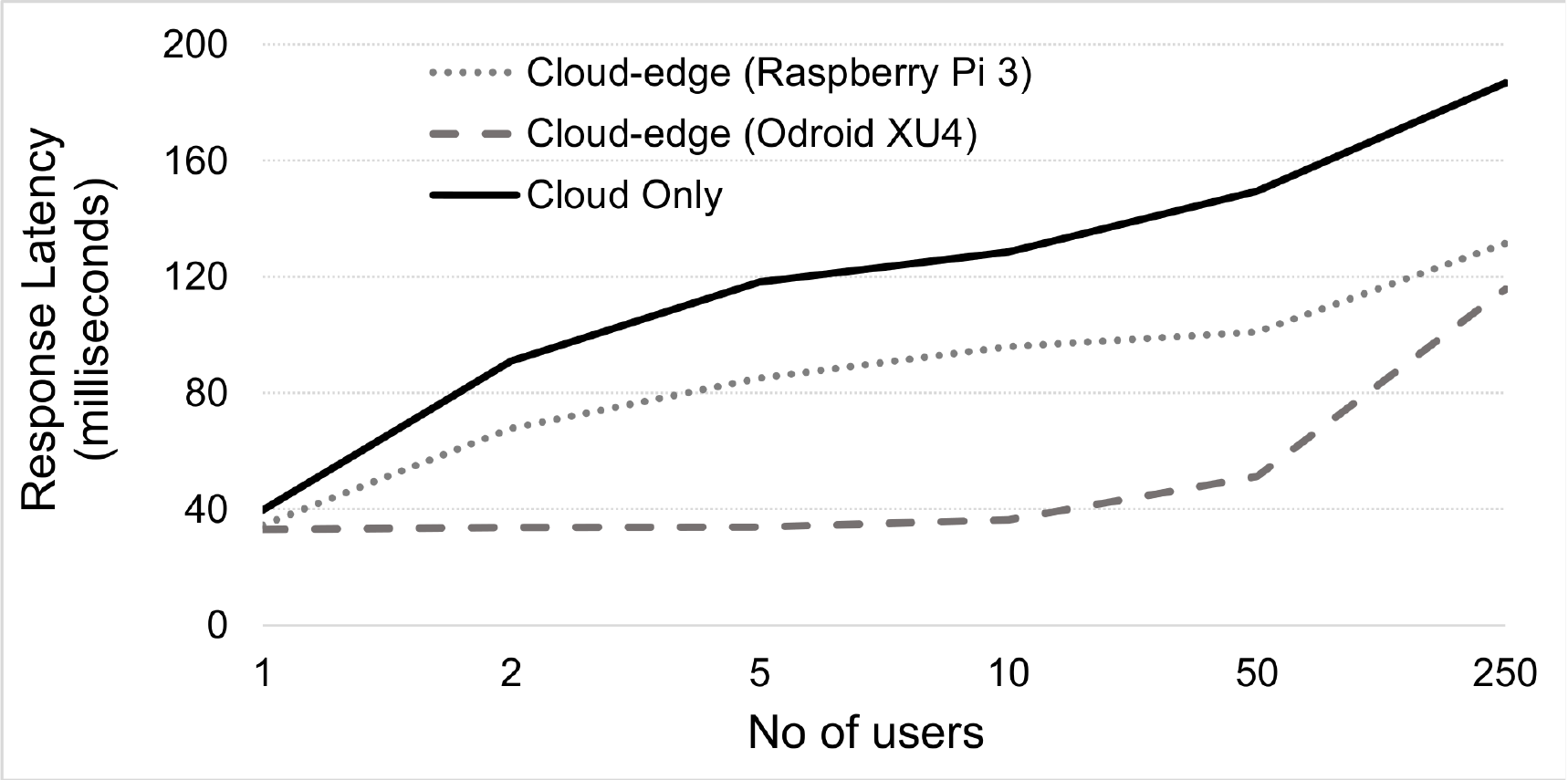}
	\caption{Impact of concurrent users on average response latency per request for iPokeMon.}
\label{fig:jmeterusers}
\end{figure}

\subsubsection{Impact of concurrent users on application latencies}
For this experiment, concurrent users (2, 5, 10, and 50) of the application benchmark are considered as shown in Figure~\ref{fig:userslatencies}. The individual requests from the application are simulated for concurrent users using JMeter. It is noted that the communication latency as shown in Figure~\ref{fig:userscommunicationodroid} and Figure~\ref{fig:userscommunicationraspberry} increases with the number of users. This is because the time in flight and time to transfer the results also increases because with the increasing number of users the rate at which data movement occurs decreases (this is highlighted in Figure~\ref{fig:usersbytestransfer}). Similar increases are noted for computation latency. Although the rate at which the computation latency increases is different. For example, FogLAMP has a much lower rate of increase in computation latency when compared to PocketSphinx. This is because of the nature of the computational intensity underpinning the benchmarks. The figures highlight the need for \texttt{DeFog} - to differentiate workloads that may not have a significant increase in computation or communication latencies even when there are multiple users versus those that may have the computation or communication latencies significantly affected.  

Applications that have a larger execution time for a single user, are impacted the most by concurrent users, which results in larger computation latencies. For example, consider PocketSphinx as shown in Figure~\ref{fig:usersrtf}. Given the current decomposition of services for the PocketSphinx application it is evident that the cloud-edge deployment is not advantageous over the cloud-only deployment. 

However, consider the impact of concurrent users on iPokeMon as shown in Figure~\ref{fig:jmeterusers}. The cloud-edge deployment clearly has significant advantages over the cloud-only deployment. Latency spike observed for the edge nodes when there are 250 concurrent users was due to a single communication request that increased the average response time of the requests. 


\subsubsection{Summary}
The experimental results have demonstrated that the communication and computation latencies vary across the different deployments, namely cloud-only, edge-only and cloud-edge. Although the communication latency of the application may improve by using the cloud-edge deployment, there will be no overall gain if the computation latency is high on resource constrained edge nodes. Applications are noted to behave differently when the edge node is stressed, both its system and network resources. The rate at which performance degrades varies across applications. Similarly, when concurrent users request service from the same edge node, the rate at which the communication and computation latencies increase vary; some workloads exhibit significant increase in latencies where as others have a negligible increase. These observations highlight the use of \texttt{DeFog} 
to identify any performance gain in using the Fog over the cloud-only deployment (\textbf{\textit{Q1}}), which services of an application would benefit from moving to the edge (\textbf{\textit{Q2}}), and which deployment platforms are best suited for an application when there are multiple hardware choices (\textbf{\textit{Q3}}).

%% file: relatedwork.tex
There are multiple techniques for understanding the performance of a target platform and applications running on it. Two dominant techniques, include simulation-based and benchmarking approaches, which are applicable in the context of edge computing~\cite{acmresmgmtsurvey-01}. 

The simulation-based approach uses computational models to simulate the target platform. The application is modelled on top of the simulated platform. The application and target environment are usually abstract representations of the real application and the underlying hardware platform on which it will run is defined by a fine-grain or aggregate set of input parameters. The accuracy of running a simulation to gather performance metrics is fully dependent on the underpinning models, which may be simply abstract representations or more fine-grain models. It is worthwhile to note that it is harder to accurately model the variability seen on a real target platform, and is more complex when modelling distributed systems, such as the Fog. Popular Fog and Edge computing based simulators include EdgeCloudSim~\cite{edgecloudsim-01}, iFogSim~\cite{ifogsim-01}, MyiFogSim~\cite{myifogsim-01} and FogExplorer\footnote{https://openfogstack.github.io/FogExplorer/}.
Recently, emulation is also used to evaluate the performance of Fog applications~\cite{mockfog-01}.

The advantages of using simulation-based approaches include reproducibility of the experiments and is a cheaper solution to understand the performance of a target platform since physical hardware is not required. However, simulation approaches may not be sufficiently accurate and running fine-grain simulation models (such as discrete event simulators) can be time consuming. 

On the other hand benchmarking approaches run the application on the target platform in more realistic conditions. Nonetheless, external factors that influence performance may be simulated; for example, network conditions or noise on the target platform. Since benchmarking is usually performed in the real setting, reproducibility of results depends on how transient the platform is. For example, reproducing results on the Fog, could be challenging. Nonetheless, benchmark consistency can be achieved by using technologies, such as containers. Currently no Fog benchmarks are readily available. 

Accuracy and reliability of results are advantages of benchmarking since applications are executed on the target platform with fewer abstractions. However, the platform needs to be set up, which is more expensive than simulation. The application source code may also need to be modified if multiple target platforms have to be benchmarked. This paper adopts a benchmarking approach. 

Existing benchmarks relevant to the discussion of this paper, include those for the cloud, end user-devices, and application specific benchmarks. Benchmarking on the cloud typically captures the performance of an application across different categories of resources, such as virtual machines (VMs)~\cite{cloudcontainerbenchmarking-01} and storage services~\cite{cloudstoragebenchmarking-01}. 
Popular cloud benchmarks include Yahoo Cloud Serving Benchmark (YCSB)~\cite{ycsb-01}, CloudRank-D~\cite{cloudrankd-01}, and DCBench~\cite{dcbench-01}.
The use of containers for benchmarking in the cloud environment has advantages, such as consistency and making the benchmarking process faster and automating it~\cite{cloudcontainerbenchmarking-02}. However, as expected cloud benchmarking methods do not need to include techniques to capture performance of resources outside the cloud and consequently, does not need to handle the more complex dependencies that are seen in Fog application. In this paper, container-based benchmarking is extended towards taking into account the dependencies between cloud-edge services of an application as seen in the Fog.

There are benchmarking methods available for end user-devices, such as mobile devices~\cite{benchmarkingleadedge-01} and IoT devices~\cite{benchmarkingiot-01}. There is recent research that focuses on understanding the interactions between the mobile user and the edge system~\cite{benchmarkingdeviceedge-01}. IoT benchmarks focus on power consumption\footnote{\url{https://www.eembc.org/products/}}, Constrained Application Protocol (CoAP) benchmarking~\cite{benchmarkingiot-01}, and low power wireless network applications (such as the IoT Benchmarking Consortium's\footnote{\url{https://www.iotbench.ethz.ch}} D-Cube~\cite{dcube-01}). Serverless application benchmarking for CPU intensive benchmarks on edge platforms, such as Amazon Greengrass and Microsoft Azure IoT Edge are considered in literature; multiple heterogeneous cloud functions are benchmarked~\cite{edgebench-01}. More application specific benchmarks have been developed. These include the TPC Express Benchmark IoT (TPCx-IoT)\footnote{\url{http://www.tpc.org/tpcx-iot/}} that is based on YCSB~\cite{ycsb-01} and considers data ingestion and query workloads. CAVBench is developed for benchmarking connected and autonomous vehicle applications~\cite{cavbench-01}. 

The general observation is that there are no Fog benchmarking suites. While there are those that specifically consider the cloud or end user-devices, benchmarking the distribution of applications across the cloud and edge has not been the focus. This paper reports a first Fog benchmarking suite. 

%% file: conclusions.tex
There are currently no readily available Fog benchmarks to address the following questions: (\textbf{\textit{Q1}}) How can the relative performance gain of using Fog platforms over the Cloud-only computing model be quantified? (\textbf{\textit{Q2}}) If there are multiple services of an application that can be moved to the edge, then how can the overall performance of moving different services to the edge be quantified? (\textbf{\textit{Q3}}) If there are multiple competing resources at the edge for a service, then which resource should be selected? \texttt{DeFog}, the benchmarking suite proposed in this paper addresses the above questions and will be a useful tool for hardware vendors, system software developers or Internet Service Providers deploying micro edge data centres.

\texttt{DeFog} comprises six benchmarks that can be deployed across the cloud and the edge. The current portfolio of application benchmarks in \texttt{DeFog} includes, deep learning-based object detection, text-to-speech conversion, text-audio forced alignment, geo-location based online mobile game, IoT edge gateway application and real-time face detection in video streams, that are deployed across a cloud-only, edge-only, and cloud-edge deployment modes. The approach taken in this paper is benchmarking actual target platforms rather than using simulation. Experimental studies demonstrate the catalogue of target platform and application related metrics captured by \texttt{DeFog}. The experiments also present the performance of benchmarks due to concurrent users and stress on edge resources. 

The approach taken by \texttt{DeFog} ensures that the build and deployment of the application are automated. \texttt{DeFog} ensures consistency (same environment is available and dependencies are set up when comparing two different edge nodes) when benchmarking.

\textbf{\textit{Limitations and Future Work}}:
The current approach in \texttt{DeFog} assumes that all target platforms will be able to run Docker containers. The container technology is limited in that processor architecture specific images are required. Relying on containers as the deployment technology for Fog computing will result in the need to host a wide range of container images for application benchmarks. More hardware agnostic deployment technologies will need to be developed to alleviate this problem. It is also assumed in the implementation and experimentation that dedicated single board computer capabilities are available at the edge. Edge resources, such as routers and gateways that can be augmented with compute capabilities have not been considered. The partitioning of applications is inherently across a single layer of cloud resources and a single layer of edge resources. The current research does not explore the distribution of benchmarks across the entire cloud-edge continuum. This is because applications that can leverage the entire cloud-edge continuum are not fully understood.  
The time taken for enacting security measures in a cloud-edge application is not considered. This will be considered in the future to quantify performance more accurately. Efforts will be made towards identifying appropriate techniques to capture or estimate the overall energy consumption in the Fog. The coverage of benchmarks will be widened in the future to include applications that can ultimately test Fog computing by simultaneously stressing all dimensions of latency, bandwidth and availability of resources, such as augmented reality applications. 

\texttt{DeFog} is open source (\url{https://github.com/qub-blesson/DeFog}) and our vision is to foster community growth and development of Fog benchmarks that can be widely used. 